\documentclass[10pt,twocolumn,twoside]{IEEEtran}
\usepackage{amsmath}
\usepackage{amssymb}

\usepackage{graphicx}
\usepackage{epstopdf}
\usepackage{amsmath}
\usepackage{array}
\newcommand{\PreserveBackslash}[1]{\let\temp=\\#1\let\\=\temp}
\newcolumntype{C}[1]{>{\PreserveBackslash\centering}p{#1}}
\newcolumntype{R}[1]{>{\PreserveBackslash\raggedleft}p{#1}}
\newcolumntype{L}[1]{>{\PreserveBackslash\raggedright}p{#1}}
\usepackage[usenames]{color}
\usepackage{colortbl,booktabs}
\usepackage{tabularx}
\usepackage{multicol}
\usepackage{booktabs}
\usepackage[Symbol]{upgreek}
\usepackage{bm}
\usepackage{cite}
\usepackage{balance}
\usepackage[linesnumbered,ruled]{algorithm2e}
\usepackage{mathrsfs}
\usepackage{url}
\usepackage{subfigure}
\usepackage{threeparttable}
\usepackage{amsmath}
\usepackage{dcolumn}
\usepackage{multirow}
\usepackage{booktabs}
\usepackage{amsfonts}
\usepackage{amsthm,amsmath,amssymb}
\usepackage{mathrsfs}
\usepackage{algorithmic}
\allowdisplaybreaks[3]
\begin{document}
	\title{Multi-Carrier Faster-Than-Nyquist Signaling for OTFS Systems\vspace{-2 mm}}
	%
	%\author{\IEEEauthorblockN{Wenqian Shen, Linglong Dai, and Zhaocheng Wang}\vspace{-0mm}}
	\author{Xueyang Wang,~\IEEEmembership{}
		Shiqi Gong,~\IEEEmembership{}
		Wenqian Shen,~\IEEEmembership{} \\
		Chengwen Xing,~\IEEEmembership{Member, IEEE},  and 
		J. Andrew Zhang,~\IEEEmembership{Senior Member, IEEE}.
		
		\vspace{-3mm}
		\thanks{	
		Xueyang Wang, Wenqian Shen, Chengwen Xing are with the School of Information and Electronics, Beijing Institute of Technology, Beijing 100081, China (E-mails: xywang1103@gmail.com; shenwq@bit.edu.cn; chengwenxing@ieee.org).
		Shiqi Gong is with the School of Cyberspace Science and
		Technology, Beijing Institute of Technology, Beijing 100081, China (e-mail:
		gsqyx@163.com).	
		J. Andrew Zhang is with the Global Big Data Technologies Centre,
		University of Technology Sydney, Sydney, NSW 2007, Australia (e-mail:
		Andrew.Zhang@uts.edu.au).}
		
		%\thanks{A. Sayeed is with the Department of Electrical and Computer Engineering, University of Wisconsin, Madison, WI 53706, USA (email: akbar@engr.wisc.edu).}
		\vspace{-5mm}
	}

	\maketitle
	\begin{abstract}
		
		%利用时域FTN和频域NOFDM可以增大传输速率，然而之前工作聚焦于AWGN信道和频选信道，并未考虑time-varying信道，然而新一代移动通信需要告诉移动，所以
		Orthogonal time frequency space (OTFS) modulation technique is promising for high-mobility applications to achieve reliable communications. However, the capacity of OTFS systems is generally limited by the Nyquist criterion, requiring orthogonal pulses in both time and frequency domains. In this paper, we propose a novel multi-carrier faster-than-Nyquist (MC-FTN) signaling scheme for OTFS systems. By adopting non-orthogonal pulses in both time and frequency domains, our scheme significantly improves the capacity of OTFS systems. Specifically, we firstly develop the signal models for both single-input single-output (SISO) and multiple-input multiple-output (MIMO) OTFS systems. Then, we optimize the delay-Doppler (DD) domain precoding matrix at the transmitter to suppress both the inter-symbol interference (ISI) and inter-carrier interference (ICI) introduced by the MC-FTN signaling. For SISO systems, we develop an eigenvalue decomposition (EVD) precoding scheme with optimal power allocation (PA) for achieving the maximum capacity. For MIMO systems, we develop a successive interference cancellation (SIC)-based precoding scheme via decomposing the capacity maximization problem into multiple sub-capacity maximization problems with largely reduced dimensions of optimization variables. Numerical results demonstrate that our proposed MC-FTN-OTFS signaling scheme achieves significantly higher capacity than traditional Nyquist-criterion-based OTFS systems. Moreover, the SIC-based precoding scheme can effectively reduce the complexity of MIMO capacity maximization, while attaining performance close to the optimal EVD-based precoding scheme.

		%Our proposed scheme provides an effective solution for future wireless communication systems in high-mobility scenarios.
		
	\end{abstract}
	
	\IEEEpeerreviewmaketitle
	
	\section{Introduction}\label{S1}
	\IEEEPARstart{}{}
	%不要公式，纯用文字描述
	%1）总体描述，FTN定义，和Nyquist比较的优势。在双选信道下应用的优势，同时OTFS信道是双选信道中的有效技术。
	%2）现有的FTN precoding，有无PA，按系统/类型分类，MC-FTN,NOFDM，
	%3）OTFS precoding较少，通常基于Nyquist而非FTN。
	
	Faster-than-Nyquist (FTN) signaling has been proposed by using non-orthogonal pulses in the time domain, where the symbol interval is shorter than that defined by the Nyquist criterion~\cite{FTN1,FTN2,FTN3,FTN4,FTN5}. Compared to the Nyquist-criterion-based signaling, FTN signaling can improve the information rate without extra bandwidth and power. Specifically, for the {\color{red} doubly-selective} fading channel in high-mobility scenarios, FTN signaling can also achieve higher capacity performance than the Nyquist-criterion-based counterpart due to the shorter symbol interval in the time domain. However, the excessive Doppler spread in high-mobility scenarios may erode the orthogonality of traditional orthogonal frequency division multiplex (OFDM) subcarriers, which makes the classical OFDM modulation technique not available for the {\color{red} doubly-selective} fading channel. As a remedy, a novel orthogonal time frequency space (OTFS) modulation technique has received much attention, which transforms the time-varying channel into a time-invariant channel in the delay-Doppler (DD) domain~\cite{OTFS1,OTFS2}. Hence, the DD domain data symbols can experience a roughly constant channel, which leads to greatly improved performance in time-varying channels. Therefore, in order to improve information rate and support stable communication in high-mobility scenarios, the integration of FTN signaling and the OTFS system is a promising technique.

%	Nyquist criterion is usually obeyed to avoid the inter symbol interference (ISI) for the signal transmission, where the minimum symbol interval is $T_0=\frac{1}{2W}$ [sec], with $2W$ Hz being the bandwidth of the bandlimited communication systems~\cite{EVD}.
%	 Hence, the achievable symbol rate is no more than $\frac{1}{T_0}$. 
%	 To improve the achievable symbol rate, faster-than-Nyquist (FTN) signaling was proposed  by using non-orthogonal pulses in the time domain with a symbol interval of $T=\tau T_0 (0<\tau\leq1)$, where $\tau$ is the symbol's packing ratio~\cite{FTN1,FTN2,FTN3,FTN4,FTN5}. 

%	The main challenge of FTN signaling is how to eliminate the FTN-induced ISI. One solution is to design detection algorithms at the receiver, which brings excess detection complexity~\cite{EQUA1,EQUA2,EQUA3}. 
	
	Precoded FTN signaling has been proposed to eliminate the effect of FTN-induced inter-symbol interference (ISI) at the transmitter~\cite{PRE1,PRE2,PRE3}. Gattami {\it et al.} utilized the inverse square root of the FTN-induced ISI matrix as the precoding scheme~\cite{PRE1}. Qian {\it et al.} proposed a singular value decomposition (SVD)-based precoding scheme for eliminating the ISI effects by diagonalizing the FTN-induced ISI matrix~\cite{PRE2}. However, these previous precoding schemes did not consider power allocation (PA) among symbols. Ishihara {\it et al.} proposed an SVD-based precoding scheme with optimal PA by maximizing the mutual information of this FTN signaling system~\cite{PRE3}. It has been shown that the capacity of the SVD precoding-based FTN system with PA is higher than that of the unprecoded FTN system. However, the precoding schemes mentioned above only considered the ideal additive white Gaussian
	noise (AWGN) channel and are only applicable to narrowband systems.
	
{\color{black}For wideband systems,
an EVD-precoding FTN signaling scheme with PA has been proposed~\cite{EVD}, and it outperforms the conventional counterparts in terms of the capacity and bit error ratio (BER) performance. In order to further improve the spectral efficiency,
 the multi-carrier FTN (MC-FTN) signaling scheme has been explored.} There are two types of MC-FTN signaling schemes, one of which adopts conventional orthogonal pulses in the frequency domain. For example, Wang {\it et al.} proposed a MC-FTN signaling scheme for frequency-selective channels and designed a linear minimum mean square error (LMMSE) equalizer to tackle the induced ISI~\cite{MFTN}. Furthermore, Ishihara {\it et al.} proposed a differential MC-FTN signaling scheme for the {\color{red} doubly-selective} fading channels~\cite{Ishihara}, where differential encoding is used in MC-FTN signaling to reduce the frame duration. 
However, these works consider adopting non-orthogonal pulses only in the time domain to improve capacity performance. In fact, the capacity performance can be further improved by adopting  non-orthogonal pulses in both time and frequency domains, which is referred to as the second type of MC-FTN signaling. For example, Dasalukunte {\it et al.} developed the hardware architecture of MC-FTN transceivers~\cite{NTF}, which are non-orthogonal in both time and frequency domains. However, this work ignored the input-output relationship and capacity analysis of the MC-FTN signaling system. Fagorusi {\it et al.} proposed the transceiver architecture of MC-FTN signaling with a specific matched filter at the receiver for tackling the MC-FTN-induced interference~\cite{Fag}. The formulation of the MC-FTN signaling in different domains also has been studied in several papers.  Ma {\it et al.} constructed a segment-based received signal model in the frequency domain and then proposed a parametric bilinear generalized approximate message passing (PBiGAMP)-based receiver for mitigating the inherent ill-conditioning problem of the MC-FTN signaling scheme~\cite{mys3}. After this work, they proposed a novel received signal model in the frequency domain for the MC-FTN signaling scheme and derived a low-complexity GAMP-based equalization algorithm~\cite{mys2}. They then proposed an index-modulated MC-FTN signaling scheme for higher spectral efficiency and derived the input-output relationship in the time domain and the frequency domain, respectively~\cite{mys}.
 These signaling schemes, which adopt non-orthogonal pulses in both time and frequency domains, are mainly for time-invariant frequency-selective channels and based on conventional multi-carrier modulation techniques such as OFDM.
 %然而下一代移动通信需要支持告诉移动，例如高铁，LEO卫星 等，而OFDM不适用。
 %OTFS被提出，介绍OTFS，再接OTFS precoding
 {\color{red}However, next-generation wireless communication systems need to support reliable communications in high-mobility scenarios, such as high-speed railways and lower earth orbit (LEO) satellites. Such high-mobility scenarios are often subject to severe Doppler shifts, rendering conventional OFDM modulation unsuitable~\cite{EXCEL,EXCEL2,OTFSgz}.
MC-FTN signaling with the reduced symbol interval and subcarrier spacing is capable of alleviating the effect of Doppler shifts caused by rapidly time-varying channels. Therefore, the MC-FTN signaling can be regarded as an effective approach to enhance communication performance in such high-mobility scenarios.}

 {\color{black}Recently, OTFS has been widely used for tackling the excessive Doppler spread caused by rapidly time-varying channels~\cite{OTFS1,OTFS2}.}
%	Therefore, OTFS is able to deal with the problems caused by the rapid time variation, such as ICI caused by the excessive Doppler spread. 
%	Another advantage of OTFS is that it can be easily implemented by adding the pre-processing and post-processing units into OFDM systems~\cite{OTFS1}. 
	%介绍OTFS precoding，需要加一句过渡,补充参考文献
	%.DD域低复杂度THP预编码，实现近似最优性能
	%.DD域多用户MIMOOTFS预编码
	%affine-precoded OTFS，将data和pilot混在一起再分开
	%DD域做data预编码，TF放置导频，可以实现精确预编码降低信道开销
	%TF域自适应预编码，
	Transmit precoding schemes for OTFS systems have also been widely investigated. However, current precoding schemes are mainly based on the Nyquist criterion. Most of the precoding schemes focus on the DD {\color{red}domain}. For example, Pandey {\it et al.} proposed a low-complexity multiuser precoder in the DD domain for the OFDM-based OTFS system~\cite{Pandey}, whose spectral efficiency performance is better than the traditional OFDM counterpart. Li {\it et al.} proposed a simple implementation of  Tomlinson-Harashima precoding (THP) in the DD domain for the OTFS system~\cite{lsy}, which attains performance close to the optimal case in the high signal-to-noise ratio (SNR) regime. Mehrotra {\it et al.} developed a new architecture that uses the affine-precoders to decouple the superimposed pilot and data symbols in the DD domain~\cite{Mehrotra}. Besides, there are also some works focusing on the precoding schemes in the frequency domain. Zhang {\it et al.} restructured the OTFS modulation as a precoded OFDM system and proposed an adaptive transmission scheme based on the precoding matrix in the frequency domain~\cite{Zhang}.  Moreover, precoding schemes can also be performed in different domains. For example, Pfadler {\it et al.} precoded data symbols in the DD domain and arranged unprecoded pilot symbols in the time-frequency (TF) domain to avoid leakage effects~\cite{Pfadler}.  Srivastava {\it et al.} proposed a low-complexity analog transceiver design by coupling the beam-alignment in the angular domain with sparse channel estimation in the DD domain~\cite{Sri}. Liu {\it et al.} proposed a low-complexity hybrid beamforming scheme for MIMO-OTFS systems by using an interleaver in the DD domain and performing the digital and analog beamforming in the time domain~\cite{Liu}. {\color{red}However, the capacity of the aforementioned OTFS systems is generally  limited by the Nyquist criterion. 
		Inspired by the idea of MC-FTN signaling, we aim to enhance the  capacity  of  doubly-selective channels by using non-orthogonal pulses in both time and frequency domains for OTFS systems. Furthermore, since MC-FTN signaling reduces the symbol interval and subcarrier spacing, the limited wireless resources can be fully exploited to  enhance OTFS modulation performance in time-varying channels.} Our work is different from the system model in \cite{EVD} where non-orthogonal pulses are exclusively utilized in the time domain and only frequency-selective channels are taken into consideration. %MIMO要写进去吗
	 Specifically, we propose a novel MC-FTN-OTFS signaling scheme to mitigate the ISI and inter-carrier interference (ICI) induced by non-orthogonality in both time and frequency domains and thus enhance the capacity in high-mobility scenarios.
	The main contributions of this paper are summarized as follows:	
	\begin{itemize}
		\item {\color{black}Motivated by the concept of the precoded FTN signaling~\cite{PRE3,EVD}, we propose a novel MC-FTN-OTFS signaling scheme for SISO and MIMO systems.
			%, which can support reliable communication and achieve higher capacity in high-mobility scenarios simultaneously.
		 More specifically, we first derive the input-output relationships of the proposed signaling schemes for SISO systems in both TF and DD domains, where the non-orthogonal pulses are employed in both time and frequency domains. We then
		 extend the MC-FTN-OTFS signaling scheme to the MIMO system and provide  the input-output relationship in a vector form.}

		\item For SISO systems, we propose an EVD precoding scheme with PA  to eliminate the MC-FTN-induced ISI and ICI. {\color{black}Specifically, we first derive the capacity of the proposed EVD-precoded MC-FTN-OTFS signaling scheme for the doubly-selective fading channels. Then we optimize the PA matrix such that the system capacity can be maximized, leading to a higher capacity than the conventional MC-FTN-OTFS signaling and Nyquist-criterion-based schemes. }

		\item For MIMO systems, motivated by the idea of successive interference cancellation (SIC)~\cite{SIC1}, {\color{black}we propose a low-complexity SIC-based precoding scheme for the MIMO MC-FTN OTFS system, in which the capacity maximization problem  is decomposed into multiple subproblems in terms of each data stream in the DD domain. After this decomposition, we resort to  solve these subproblems by utilizing the proposed EVD
		precoding scheme with PA, where the optimal  precoding matrices associated with  different data streams  can be successively obtained.}

	\end{itemize}

	Numerical results demonstrate that the capacity performance of our proposed EVD precoded MC-FTN-OTFS signaling scheme with optimal PA outperforms the classical Nyquist-criterion-based signaling scheme and FTN signaling scheme in terms of capacity but experiences a slight reduction in BER performance. 
	 Moreover, the SIC-based precoding method effectively avoids the high computational burden of the capacity maximization problem at the cost of the  minimal loss of capacity.

	The rest of this paper is organized as follows. In Section~\ref{S2}, we present signal models of the MC-FTN-OTFS modulation scheme for both SISO and MIMO systems and then derive the input-output relationships of the proposed scheme. In Section III, we propose an EVD-based precoding scheme with optimal PA to maximize the system capacity for SISO systems. In Section IV, we develop a low-complexity SIC-based precoding scheme for MIMO systems. Finally, we provide numerical results in terms of capacity and BER performance in Section V and give our conclusions in Section VI.

	\begin{figure*}[!t] \vspace{-3mm}
		\centering
		\subfigure[Block diagram of the proposed MC-FTN-OTFS modulation/demodulation.]
		{       \includegraphics[width=18cm]{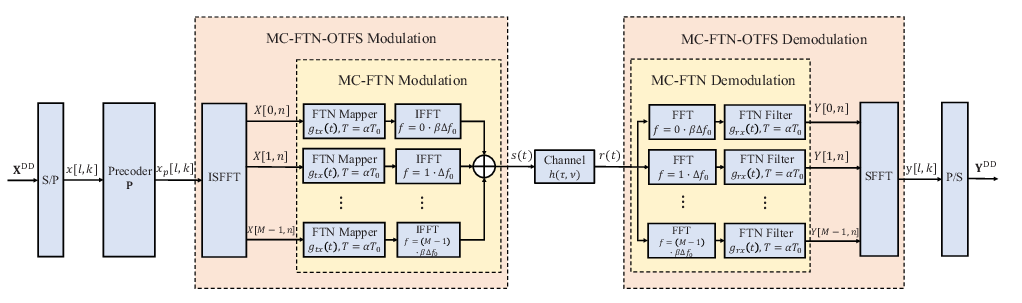}}
		\vspace{-2mm}
		\hspace{1in}
		\subfigure[Illustration of the proposed MC-FTN-OTFS signaling in both time and frequency domains.]{
			\includegraphics[width=13cm]{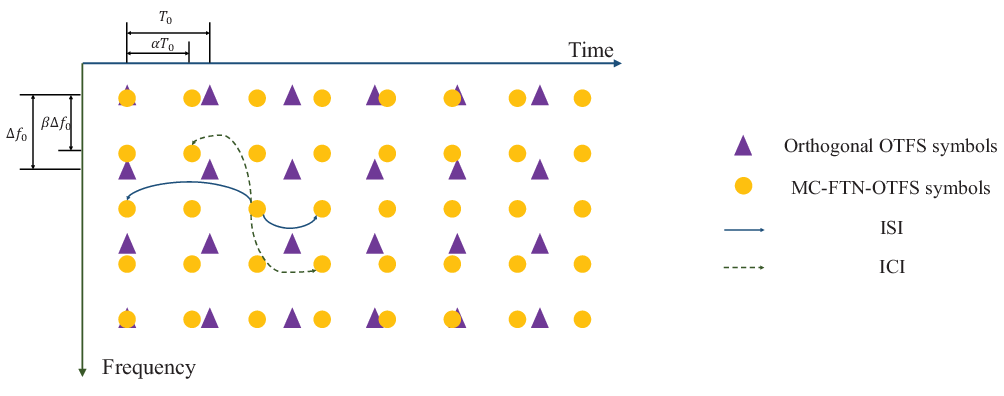}}
		\caption{The proposed MC-FTN-OTFS signaling with non-orthogonal pulses  adopted in both time and frequency domains.} 
		\vspace{-5mm}
		\label{diagram1}
	\end{figure*}

	%Our simulations verify the analytical results.~
	\emph{Notations}: We use the following notations throughout this paper. Let a, $\mathbf{a}$ and  $\mathbf{A}$ represent the scalar, the vector, and the matrix, respectively; 
	$(\cdot)^{{T}}$,
	$(\cdot)^{{H}}$, $(\cdot)^{*}$ and $(\cdot)^{-1}$ denote the transpose,
	conjugate transpose, conjugate and inverse of $(\cdot)$, respectively; $\delta( \cdot )$ is the Dirac delta function; 
	%$[\cdot]_N$ denotes modulo $N$ operation; $\lceil {a} \rceil$ represents the smallest integer no less than $a$; $\lfloor {a} \rfloor$ is the lagest integer no greater than $a$; 
	$\mathbf{F}_N$ denotes the Fourier transform matrix of size $N\times N$; $\mathbf {I}_N$ denotes an identity matrix of size $N\times N$.

	%$\rm vec\{\cdot \}$ denotes the vectorization of a matrix and $\rm ivec\{\cdot \}$ denotes the invectorization of a vector;$\left[\mathbf{a}\right]_{m:n}$ denotes the $m$-th to $n$-th element of vector $\mathbf{a}$; $\left[\mathbf{A}\right]_{m,n}$ denotes the $(m,n)$-th element of matrix $\mathbf{A}$; $\left[\mathbf{A}\right]_{:,n}$ denotes the $n$-th column of matrix $\mathbf{A}$; The operator $ \circ $, $\otimes$ and $*$ represent the Hadamard-product, Kronecker product and convolution, respectively; , and ${\mathbf{1}}_{m,n}$ denotes all 1 matrix of size $m \times n$

	\section{MC-FTN-OTFS Modulation Scheme}\label{S2}
	
	In this section, we formulate the signal models of the MC-FTN-OTFS signaling scheme for both SISO and point-to-point MIMO systems. The input-output relationships in a vector form for both systems are also derived.

	\subsection{SISO Signal Models}
	
	As illustrated in Fig. 1(a), we consider an OTFS system with MC-FTN signaling, which is also referred to as the MC-FTN-OTFS system. Let $M$ and $N$ denote the numbers of delay and Doppler bins, respectively. We assume $\mathbf{X}^\mathrm{DD} \in \mathbb{C}^{M\times N}$ to be the delay-Doppler (DD) domain information symbols, whose $(l,k)$-th element $x[l,k]$ is a complex-valued Gaussian symbol with the average  power $E_x = \mathbb{E}[\left|x[l,k]\right|^2] = \sigma_x^2$ ,$l=0,\cdots,M-1$, $k=0,\cdots,N-1$.
	%In this section, we formulate the system model of ultra-dense FTN signaling for OTFS system, where the time interval is lower than the minimum Nyquist time interval and the subcarrier spacing is lower than the minumum orthogonal subcarrier spacing. 
	The symbol block $\mathbf{X}^\mathrm{DD}$ is firstly vetorized as $\mathbf{x}^\mathrm{DD}\in \mathbb{C}^{MN\times 1}$ and then precoded by a DD domain linear precoding matrix $\mathbf{P}\in \mathbb{C}^{MN \times MN}$. Therefore, the precoded DD symbols  can be  expressed as
	\begin{align}
	\label{x-precode} \mathbf{x}^\mathrm{DD}_P=\mathbf{P}\mathbf{x}^\mathrm{DD},
	\end{align}
	where  the $(lM+k+1)$-th element of the vector $\mathbf{x}^\mathrm{DD}_P\in \mathbb{C}^{MN\times 1}$ is $x_P[l,k]$. Design of the precoder will be detailed in Sections \ref{S3} and \ref{S4}.
	Based on the inverse symplectic finite Fourier transform (ISFFT), the proposed MC-FTN-OTFS maps $x_P[l,k]$ to TF symbols $X_P[m,n]$ as follows~\cite{Inteference,OTFS1,OTFS2}:
	\begin{align}\label{XTF}
	X_P[m,n]=\frac{1}{\sqrt{N M}} \sum_{k=0}^{N-1} \sum_{l=0}^{M-1} x_P[l,k] e^{j 2 \pi\left(\frac{n k}{N}-\frac{m l}{M}\right)}, \notag \\\quad m=0,\ldots,M-1, n=0,\ldots,N-1.
	\end{align}
	
	%Vectorizing $\mathbf{X}^\mathrm{TF}$, we have
	%\begin{align}\label{xTF}
	%\mathbf{x}^\mathrm{TF} = \operatorname{vec}(\mathbf{X}^\mathrm{TF} ) = \left(\mathbf{F}_N^{\mathrm{H}} \otimes \mathbf{F}_M\right) \mathbf{x}^\mathrm{dD}_P.
	%\end{align}
	
	Next, the TF symbols $X_P[m,n]$'s are then processed by the proposed MC-FTN modulator. Specifically, the set of TF symbols $X_P[m,n]$ are firstly transformed into $M$ parallel subcarriers, each of which consists of $N$ symbols and then passes through the conventional FTN mapper, where the symbol interval $T=\alpha T_0  (0< \alpha \leq 1)$ is shorter than that defined under ISI-free Nyquist criterion $T_0$ and $\alpha$ is the  compression factor in the time domain. The FTN mapper outputs pass through the IFFT module, where the subcarrier spacing $\Delta f = \beta \Delta f_0 (0< \beta \leq 1)$ is lower than that required in the classical OFDM $\Delta f_0$ and $\beta$ is the compression factor in the frequency domain. Generally, we have $T_0 \Delta f_0=1$\cite{OTFS1}.  
	
	Symbols in time and frequency domains are arranged as Fig.1 (b) shows. In the conventional OTFS system, the time interval is $T_0$ and the subcarrier spacing is $\Delta f_0 = \frac{1}{T_0}$. Whereas for the proposed MC-FTN-OTFS system, the intervals are folded in both time and frequency domains to $T = \alpha T_0$ and $\Delta f = \beta \Delta f_0$, which causes ISI and ICI.
	
	Then the modulated MC-FTN signals $s(t)$ in the time domain is given by
	\begin{align}
	\label{h-prior} s(t)= {\sqrt{\alpha \beta E_0}} \sum_{m=0}^{M-1}\sum_{n=0}^{N-1}X_P[m,n] g_{tx}(t-n\alpha T_0) \notag \\ e^{j2\pi m\beta \Delta f_0(t-n\alpha T_0)} ,
	\end{align}
	where $g_{tx}(t)$ is the impulse response of a band-limited root raised-cosine (RRC) pulse shaping filter with a roll-off factor $\theta$, which is different from the ideal bi-orthogonal and practical rectangular pulses adopted in the conventional OTFS system. {Moreover, $E_0$ is the transmit power per subcarrier in the classical OFDM in the TF domain, and
		${\sqrt{\alpha \beta E_0}}$ is the power normalization factor.} 
	
	Assume that the time-varying channel propagation consists of $L$ paths. Let $h_i$, $\tau_{i}$ and $\nu_{i}$ denote the channel gain, delay and Doppler of the $i$-th path, respectively. The channel response in the {\color{red}DD domain} can be expressed as
	\begin{align} \label{hdD}
	h(\tau, \nu)=\sum_{i=1}^{L} h_{i} \delta\left(\tau-\tau_{i}\right) \delta\left(\nu-\nu_{i}\right).
	\end{align}
	
	Then the received signal $r(t)$ after passing through the time-varying channel is given by 
	\begin{align} \label{rt}
	r(t)=\iint h(\tau, \nu) s(t-\tau) e^{j 2 \pi \nu(t-\tau)} d \tau d \nu + z_n(t),
	\end{align}
	where  {\color{black}$z_n(t)$ is the Gaussian noise with zero mean and variance $N_0$.} 
	
	Based on the MC-FTN demodulation at the receiver, a matched filter $g_{rx}(t)=g^*_{tx}(-t)$ is applied to the received signal and then the  processed signal is  projected onto the non-orthogonal subcarriers. 
	Hence, the received TF signal $Y(f,t)$ is calculated by the cross-ambiguity function $A_{g_{rx},r}(f,t)$, which can be represented as
	\begin{align}\label{Ytf}
	Y(f,t)=A_{g_{rx},r}(f,t)=\int g_{\mathrm{rx}}^{*}\left(t^{\prime}-t\right) r\left(t^{\prime}\right) e^{-j 2 \pi f\left(t^{\prime}-t\right)} d t^{\prime}.
	\end{align}
	We can then obtain the matched filter output after sampling at the frequency $f=m\beta\Delta f_0$ and the time $t=n\alpha T_0$, which is given by
	\begin{align}\label{Ynm}
	Y[m,n]=\left.Y(f,t)\right|_{f=m\beta  \Delta f_0,  t=n\alpha T_0}.
	\end{align}

	For the ease of analysis, we then analyze the input-output relationship between $Y[m,n]$ and $X[m,n]$ in the following  proposition~\cite{Inteference}\cite{RRC}.
	
	{\emph{Proposition 1}: 
		The input-output relationship of the MC-FTN-OTFS system in the TF domain is given by
		\begin{align} \label{iotf}
		Y[m,n]=\sum_{m^{\prime}=0}^{N-1} \sum_{n^{\prime}=0}^{M-1} H_{m,n}\left[m^{\prime}, n^{\prime}\right] X_P\left[m^{\prime}, n^{\prime}\right]+Z[m,n],\notag \\
		m=0,\ldots,M-1,n=0,\ldots,N-1,
		\end{align}	
		where
		\begin{align}
		\begin{aligned}
		H_{m,n}\left[m^{\prime}, n^{\prime}\right] \!\!
		&=\!\! \sum_{i=1}^L h_i A_{g_{\mathrm{rx}}, g_{\mathrm{tx}}}\left(\left(m-m^{\prime}\right) \beta \Delta f_0-\nu_i,  \left(n-n^{\prime}\right) \right.  \notag \\ & \!\!\!\!\!\!\!\!\!\!\!\!\!\!\! \left.  \cdot \alpha T_0-\tau_i\right)  e^{j 2 \pi\left(\nu_i+m^{\prime}\beta \Delta f_0\right)\left(\left(n-n^{\prime}\right) \alpha T_0-\tau_i\right)} e^{j 2 \pi \nu_i n^{\prime} \alpha T} ,
		\end{aligned}
		\end{align}	
		and $Z[m,n]$ $(m=0,\ldots,M-1, n=0,\ldots,N-1)$ is formulated as
		\begin{align}\label{Zmn}
		Z[m,n] & = \left.Z(f,t)\right|_{f=m\beta  \Delta f_0,  t=n\alpha T_0} \notag \\
		&=\int g_{\mathrm{rx}}^{*}\left(t^{\prime}-t\right) z_n\left(t^{\prime}\right) e^{-j 2 \pi f\left(t^{\prime}-t\right)} d t^{\prime} .
		\end{align}

		{\emph{Proof:}} The proof is given in Appendix A.
		
		Finally, by applying the SFFT on  $Y[m,n]$, we can recover the DD domain information symbols $y[l,k]$  as:
		\begin{align}\label{ykl}
		y[l,k]=\frac{1}{\sqrt{N M}} \sum_{n=0}^{N-1} \sum_{m=0}^{M-1} Y[m,n] e^{-j 2 \pi\left(\frac{n k}{N}-\frac{m l}{M}\right)}.
		\end{align}
		The derivations from (\ref{x-precode}) to (\ref{ykl})  provide the basic signal model of the  MC-FTN-OTFS system.
		
		Furthermore, we also capture the input-output relationship between the output $y[l,k]$ and the input $x[l,k]$ in the DD domain in the following  proposition.
		
		{\emph{Proposition 2}: 
			The input-output relationship of the MC-FTN-OTFS system in the DD domain is given by
			\begin{align} \label{ioDD}
			y[l,k]=\frac{1}{N M} \sum_{k^{\prime}=0}^{N-1} \sum_{l^{\prime}=0}^{M-1} h_{l,k}\left[l^{\prime}, k^{\prime}\right] x_P\left[l^{\prime}, k^{\prime}\right]+z[ l,k],
			\end{align}	
			where
			\begin{align}
			z[l,k]=\frac{1}{\sqrt{N M}} \sum_{n=0}^{N-1} \sum_{m=0}^{M-1} Z[m,n] e^{-j 2 \pi\left(\frac{n k}{N}-\frac{m l}{M}\right)}
			\end{align}	
			and 
			\begin{align}
			\begin{aligned}
			h_{l, k}\left[l^{\prime}, k^{\prime}\right]=  \sum_{n=0}^{N-1} \sum_{m=0}^{M-1} \sum_{n^{\prime}=0}^{N-1} \sum_{m^{\prime}=0}^{M-1} H_{m,n}\left[m^{\prime}, n^{\prime}\right] e^{-j 2 \pi\left(\frac{n k}{N}-\frac{m \ell}{M}\right)}   \\ e^{j 2 \pi\left(\frac{n^{\prime} k^{\prime}}{N}-\frac{m^{\prime} \ell^{\prime}}{M}\right)} .
			\end{aligned}
			\end{align}
			
			{\emph{Proof:}} The proof is given in Appendix B.
			
			{\emph{Remark 1}: 
				The compression factors $\alpha$ and $\beta$ affect the degrees of folding in the time and frequency domains, respectively. If the frequency compression factor $\beta = 1$ holds, then both (\ref{iotf}) and (\ref{ioDD}) reduce to  conventional FTN signaling for OTFS systems, where non-orthogonal pulses are only adopted in the time domain.
				Furthermore, when $\alpha = \beta = 1$ holds, the pulses in both time and frequency domains are orthogonal, which is referred to as the Nyquist-criterion-based case.
				Compared to the conventional OTFS system, the MC-FTN-OTFS system  compresses the signals in both the time and frequency domain. Therefore, with the same time duration and bandwidth, the MC-FTN-OTFS system can achieve a higher capacity.
				
				%和OFDM很难有关系。。
				
				By stacking $y[l,k]$, $x_P[l,k]$ and $z[l,k]$ for $l=0,\ldots,M-1, k=0,\ldots,N-1$ into vectors $\mathbf{y}^\mathrm{DD} \in \mathbb{C}^{MN\times1}$,  $\mathbf{x}_P^\mathrm{DD} \in \mathbb{C}^{MN\times1}$ and $\mathbf{z}^\mathrm{DD} \in \mathbb{C}^{MN\times1}$, respectively, the input-output relationship in (\ref{ioDD}) can be rewritten in a vector form as
				\begin{align} \label{ydD}
				\mathbf{y}^\mathrm{DD} = \mathbf{H}^\mathrm{DD} \mathbf{x}_P^\mathrm{DD} +\mathbf{z}^\mathrm{DD},
				\end{align}
				where $\mathbf{H}^\mathrm{DD}$ is denoted as
				\begin{align}
				\mathbf{H}^\mathrm{DD}\!\! = \!\!\!
				\left[\begin{array}{ccc}
				h_{0,0}[0,0] & \ldots & h_{0,0}[N-1,M-1]\\
				\vdots& \ddots &\vdots \\
				h_{N-1,M-1}[0,0] & \ldots & h_{N-1,M-1}[N-1,M-1]
				\end{array}\right]\!\!\! 
				\end{align}
				and $\mathbf{z}^\mathrm{DD}$ is the correlated noise, whose correlation matrix will be calculated in \ref{S3}. A.
				
				In the MC-FTN-OTFS system, symbols are multiplexed in the DD domain, where the channel is  sparse as compared to the time-varying channel in the TF domain. This characteristic allows for more accurate channel estimation~\cite{OTFSshen}, which is leveraged in the MC-FTN precoding schemes when the channel state information (CSI) is unknown, thereby enhancing overall performance. However, this aspect is beyond the scope of this paper and will be investigated in our future work.

				\subsection{MIMO Signal Models}
				
					\begin{figure}[t]\vspace{-2mm}
					\center{\includegraphics[width=1\columnwidth]{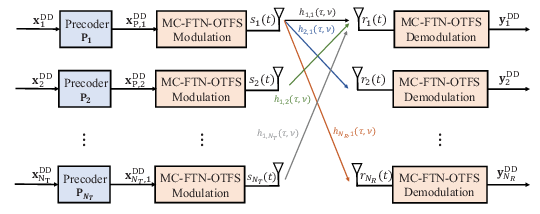}}
					\vspace{-3mm}
					\caption{The diagram of the MIMO MC-FTN-OTFS system.}
					\label{diagram2}
					\vspace{-3mm}
				\end{figure}
				
				As illustrated in Fig. \ref{diagram2}, we consider a point-to-point MIMO MC-FTN-OTFS system, where the transmitter equipped with $N_T$ antennas intends to transmit $N_T$ DD domain data streams to the receiver equipped with $N_R$ antennas and each antenna transmits OTFS modulated information symbols independently. At the transmitter, $N_T$ parallel data streams $\mathbf{x}^\mathrm{DD}_{n_t}\in \mathbb{C}^{MN\times 1}$, $n_t=1,\ldots,N_T$ are transmitted independently at $N_T$ transmit antennas. Specifically, each DD domain data stream $\mathbf{x}^\mathrm{DD}_{n_t}$ passes through a precoder $\mathbf{P}_{n_t}\in \mathbb{C}^{MN\times MN}$ and the resultant transmitted symbol at the $n_t$-th transmit antenna is given by $\mathbf{x}_{\mathrm{P},n_t}=\mathbf{P}_{n_t}\mathbf{x}^\mathrm{DD}_{n_t}$. Recall the proposed MC-FTN-OTFS modulation and demodulation schemes in Fig. \ref{diagram1} and the input-output relationship of the proposed MC-FTN-OTFS system, the DD domain received signal $\mathbf{y}_{n_r}^\mathrm{DD}\in \mathbb{C}^{MN\times 1}$ at the $n_r$-th antenna is further rewritten as
				\begin{align}\label{ynr}
				\mathbf{y}_{n_r}^\mathrm{DD} = \sum_{n_t=1}^{N_T}{\mathbf{H}_{n_r,n_t}^\mathrm{DD}\mathbf{x}_{\mathrm{P},n_t}^\mathrm{DD}}+\mathbf{z}_{n_r}^\mathrm{DD},
				\end{align}
				where  $\mathbf{H}_{n_r,n_t}^\mathrm{DD}\in \mathbb{C}^{MN\times MN}$ denotes the equivalent DD channel between the $n_t$-th transmit antenna and $n_r$-th receive antenna, in which the $(n_r,n_t)$-th element $h_{n_r,n_t}(\tau, \nu)$ is modeled similarly to (\ref{rt}) and $\mathbf{z}_{n_r}^\mathrm{DD}\in \mathbb{C}^{MN\times 1}$ is the correlated noise at the $n_r$-th antenna, which is defined similarly to $\mathbf{z}^\mathrm{DD}$ introduced in (\ref{ydD}).
				
				By stacking DD domain receive signals and DD domain noise, i.e. $\mathbf{y}_{n_r}^\mathrm{DD}$ and $\mathbf{z}_{n_r}^\mathrm{DD}$, $n_r=1,\ldots,N_R$ and  DD transmit signals $\mathbf{x}_{\mathrm{P},n_t}^\mathrm{DD}$, $n_t=1,\ldots,N_T$ into the vectors $\mathbf{y}^\mathrm{DD}_\mathrm{MIMO}\in \mathbb{C}^{MNN_R\times 1}$, $\mathbf{z}^\mathrm{DD}_\mathrm{MIMO}\in \mathbb{C}^{MNN_R\times 1}$ and $\mathbf{x}_\mathrm{P,MIMO}^\mathrm{DD}\in \mathbb{C}^{MNN_T\times 1}$, respectively, we have the following input-output relationship of the MIMO MC-FTN-OTFS system.
				
				{\emph{Proposition 3}: 
					The input-output relationship of the MIMO MC-FTN-OTFS system in the DD domain is given by
					\begin{align} \label{yDDMIMO}
					\mathbf{y}^\mathrm{DD}_\mathrm{MIMO} = \mathbf{H}^\mathrm{DD}_\mathrm{MIMO} \mathbf{P}_\mathrm{MIMO}\mathbf{x}^\mathrm{DD}_\mathrm{MIMO} +\mathbf{z}_\mathrm{MIMO}^\mathrm{DD},
					\end{align}	
					where
					\begin{align}
					\mathbf{H}_\mathrm{MIMO}^\mathrm{DD} & = 
					\left[\begin{array}{cccc}
					\mathbf{H}^\mathrm{DD}_{1,1} & \mathbf{H}^\mathrm{DD}_{1,2} & \ldots &\mathbf{H}^\mathrm{DD}_{1,N_T}\\
					\vdots&\vdots& \ddots &\vdots \\
					\mathbf{H}^\mathrm{DD}_{N_R,1} & \mathbf{H}^\mathrm{DD}_{N_R,2} & \ldots &\mathbf{H}^\mathrm{DD}_{N_R,N_T}
					\end{array}\right] \notag\\
					&\in \mathbb{C}^{MNN_R\times MNN_T}
					\end{align}
					and $\mathbf{P}_\mathrm{MIMO} = \operatorname{diag}\{\mathbf{P}_{1},\ldots, \mathbf{P}_{N_T} \} \in \mathbb{C}^{MNN_T\times MNN_T}$ is a block diagonal matrix.
					
					{\emph{Proof:}} The proof is given in Appendix C.
				
				Based on (\ref{ydD}) and (\ref{yDDMIMO}), we aim to optimize the transmit precoders for the proposed MC-FTN-OTFS SISO and MIMO systems, as will be elaborated in Sections \ref{S3} and \ref{S4}.
%				 in the following sections, respectively. Specifically, we propose an EVD-precoding scheme for SISO systems to eliminate both the MC-FTN-induced ISI and ICI, and then design the PA matrix to  maximize system capacity. Moreover, we  propose a low-complexity SIC-based precoding scheme for MIMO systems, being capable  of  achieving a capacity close to the optimal EVD-based counterpart.
				
				\section{EVD-Precoded SISO MC-FTN-OTFS Systems}\label{S3}
				In this section, we propose an EVD-based precoding scheme with optimal PA for the SISO MC-FTN-OTFS system. Specifically, we firstly derive the capacity of the EVD-precoded MC-FTN-OTFS system and then optimize the PA matrix to maximize the system capacity. 
				
				\subsection{Capacity of EVD-Precoded MC-FTN-OTFS Signaling }
				
				In this subsection, we firstly derive the mutual information of the proposed MC-FTN-OTFS signaling. 
				
			To be specific, based on the Shannon information theory, it follows from (\ref{ydD}) that the mutual information between $\mathbf{y}^\mathrm{DD}$ and $\mathbf{x}^\mathrm{DD}$ is {\color{red}given} by
				\begin{align}
				I(\mathbf{x}^\mathrm{DD} ; \mathbf{y}^\mathrm{DD}) &= h_{\mathrm{e}}(\mathbf{y}^\mathrm{DD})-h_{\mathrm{e}}(\mathbf{y}^\mathrm{DD}|\mathbf{x}^\mathrm{DD})\notag \\
				&=h_{\mathrm{e}}(\mathbf{y}^\mathrm{DD})-h_{\mathrm{e}}(\mathbf{z}^\mathrm{DD}),
				\end{align}
				where $h_{\mathrm{e}}(\cdot)$ denotes the differential entropy. The differential entropy of $\mathbf{y}^\mathrm{DD}$ is calculated as \cite{information}
				\begin{align} \label{hey}
				h_{\mathrm{e}}(\mathbf{y}^\mathrm{DD}) \leq \log _2\left((\pi e)^N \operatorname{{det}}\left({\mathbb{E}\left[\mathbf{y}^\mathrm{DD}{\mathbf{y}^\mathrm{DD}}^H\right]}\right)\right),
				\end{align}
				where the correlation matrix
				$\mathbb{E}\left[\mathbf{y}^\mathrm{DD}{\mathbf{y}^\mathrm{DD}}^H\right]$ is given by
				\begin{align} \label{corr}
				&\mathbb{E}\left[\mathbf{y}^\mathrm{DD}{\mathbf{y}^\mathrm{DD}}^H\right]\notag \\&=\mathbb{E}\left[(\mathbf{H}^\mathrm{DD} \mathbf{x}_P^\mathrm{DD} +\mathbf{z}^\mathrm{DD})(\mathbf{H}^\mathrm{DD} \mathbf{x}_P^\mathrm{DD} +\mathbf{z}^\mathrm{DD})^H\right]\notag \\ 
				&=\mathbf{H}^\mathrm{DD}\mathbf{P}\mathbb{E}\left[\mathbf{x}^\mathrm{DD}{\mathbf{x}^\mathrm{DD}}^H\right]\mathbf{P}^H{\mathbf{H}^\mathrm{DD}}^H +\mathbb{E}\left[\mathbf{z}^\mathrm{DD}{\mathbf{z}^\mathrm{DD}}^H\right] ,
				\end{align} 
				where we have $\mathbb{E}\left[\mathbf{x}^\mathrm{DD}{\mathbf{z}^\mathrm{DD}}^H\right]=\mathbb{E}\left[\mathbf{z}^\mathrm{DD}{\mathbf{x}^\mathrm{DD}}^H\right]=0$.
				For the MC-FTN-OTFS system, in order to calculate the correlation matrix $\mathbb{E}\left[\mathbf{z}^\mathrm{DD}{\mathbf{z}^\mathrm{DD}}^H\right]$, we firstly calculate the correlation of noise samples $Z[m,n]$ in the TF domain as (\ref{corr18}), which is shown at the top of the next page.
				\begin{figure*}[!h]
					\centering
				\begin{align} \label{corr18}
				\mathbb{E}\left[Z[m_1,n_1]Z^*[m_2,n_2]\right]  &= N_0 \int{g_{rx}^*(t-(n_1-n_2)\alpha T_0)g_{rx}(t)e^{-j2\pi (m_1-m_2)\beta \Delta f_0 (t-(n_1-n_2)\alpha T_0)}} dt e^{j2\pi m_2\beta \Delta f_0(n_1-n_2)\alpha T_0}\notag \\
				&=N_0A_{g_{rx},g_{rx}}((m_1-m_2)\beta \Delta f_0,(n_1-n_2)\alpha T_0)e^{j2\pi m_2\beta \Delta f_0(n_1-n_2)\alpha T_0}.
				\end{align}
				\hrulefill
				\vspace{-6mm}
				\end{figure*}
				{\color{red} Based on the ISFFT, we have $\mathbf{z}^{\mathrm{DD}} = \left(\mathbf{F}_N \otimes \mathbf{F}_M^H\right) \mathbf{z}^{\mathrm{TF}}$, where $\mathbf{z}^{\mathrm{TF}}\in \mathbb{C}^{MN\times 1}$ is constructed by stacking $Z[m,n]$ for all $m=0,\ldots,M-1, n=0,\ldots,N-1$. Therefore, the correlation matrix $\mathbb{E}\left[\mathbf{z}^\mathrm{DD}{\mathbf{z}^\mathrm{DD}}^H\right]$ can be represented in a matrix form as
				\begin{align}\label{zcorr}
				\mathbb{E}\left[\mathbf{z}^\mathrm{DD}{\mathbf{z}^\mathrm{DD}}^H\right] = N_0(\mathbf{F}_N\otimes\mathbf{F}_M^H)\mathbf{G}(\mathbf{F}_N^H\otimes\mathbf{F}_M),
				\end{align}
				where $\mathbf{G}\in \mathbb{C}^{MN\times MN}$ is referred to as the MC-FTN-induced ISI and ICI matrix~\cite{PRE3,EVD}, whose $(m_1M+n_1,m_2M+n_2)$-th element is given by  $\frac{1}{N_0}\mathbb{E}\left[Z[m_1,n_1]Z^*[m_2,n_2]\right]$.}
				
				Based on (\ref{zcorr}) and $\mathbb{E}\left[\mathbf{x}^\mathrm{DD}{\mathbf{x}^\mathrm{DD}}^H\right]=\sigma_{x}^2\mathbf{I}_{MN}$, we can further rewrite (\ref{corr}) as
				\begin{align}
				\mathbb{E}\left[\mathbf{y}^\mathrm{DD}{\mathbf{y}^\mathrm{DD}}^H\right]=
				\sigma_{x}^2 \mathbf{H}^\mathrm{DD}\mathbf{P}\mathbf{P}^H{\mathbf{H}^\mathrm{DD}}^H \notag \\+N_0(\mathbf{F}_N\otimes\mathbf{F}_M^H)\mathbf{G}(\mathbf{F}_N^H\otimes\mathbf{F}_M),
				\end{align} 
				
				Similarly, the differential entropy of the Gaussian noise vector $\mathbf{z}^\mathrm{DD}$ is given by
				\begin{align} \label{hez}
				h_{\mathrm{e}}(\mathbf{z}^\mathrm{DD}) & = \log _2\left((\pi e)^N\operatorname{{det}}\left(\mathbb{E}\left[\mathbf{z}^\mathrm{DD}{\mathbf{z}^\mathrm{DD}}^H\right]\right)\right)\notag \\
				&=\log _2\left((\pi e)^N\!\operatorname{{det}} \! \left(N_0(\mathbf{F}_N\otimes\mathbf{F}_M^H)\mathbf{G}(\mathbf{F}_N^H\otimes\mathbf{F}_M)\right)\right)\!.
				\end{align} 
				By combining (\ref{hey}) and (\ref{hez}), the mutual information between $\mathbf{y}^\mathrm{DD}$ and $\mathbf{x}^\mathrm{DD}$ is upper-bounded as (\ref{infor}), which is shown at the top of the next page.
				\begin{figure*}[!h]
					\centering
					\vspace{0mm}
				\begin{align} \label{infor}
				I(\mathbf{x}^\mathrm{DD} ; \mathbf{y}^\mathrm{DD}) &\leq \log _2 \frac{\operatorname{det}\left({\sigma_{x}^2 \mathbf{H}^\mathrm{DD}\mathbf{P}\mathbf{P}^H{\mathbf{H}^\mathrm{DD}}^H +N_0(\mathbf{F}_N\otimes\mathbf{F}_M^H)\mathbf{G}(\mathbf{F}_N^H\otimes\mathbf{F}_M)}\right)}{\operatorname{det}\left(N_0(\mathbf{F}_N\otimes\mathbf{F}_M^H)\mathbf{G}(\mathbf{F}_N^H\otimes\mathbf{F}_M)\right)}\notag\\
				&=\log _{2}\operatorname{det}\left(\mathbf{I}_{MN}+\frac{\sigma_x^2}{N_{0}} (\mathbf{F}_N^H\otimes\mathbf{F}_M)\mathbf{H}^\mathrm{DD}\mathbf{P}\mathbf{P}^H{\mathbf{H}^\mathrm{DD}}^H (\mathbf{F}_N\otimes\mathbf{F}_M^H)\mathbf{G}^{-1}\right).
				\end{align}
				\vspace{-7mm}
					\hrulefill
				\end{figure*}
				%where we have the compression factor $\alpha$ satisfied the range of $\alpha \geq 1/(1+\theta)$, which follows from \cite{rolloff}. 
				
				In (\ref{infor}), for a finite frame size $N$ in the time domain, the MC-FTN-induced ISI and ICI matrix $\mathbf{G}$ is positive definite with all eigenvalues being non-zero~\cite{EVD,rolloff}. However, some eigenvalues of $\mathbf{G}$ may be very small when $\alpha \textless 1/(1+\theta)$, and then the symbols associated with these small eigenvalues need to be deactivated to guarantee the effectiveness of the proposed signaling scheme~\cite{PRE3}. In particular, when the time frame size $N$ is infinite, some  eigenvalues of $\mathbf{G}$ become zero when $\alpha \textless 1/(1+\theta)$, thereby leading to   a  singular  $\mathbf{G}$. For the sake of simplicity, we assume $\alpha \geq 1/(1+\theta)$ to guarantee that $\mathbf{G}$ is an invertible matrix for all values of $N$~\cite{rolloff}. 
				
				Based on $\log_{2}\operatorname{det}(\mathbf{I}+\mathbf{AB})=\log_{2}\operatorname{det}(\mathbf{I}+\mathbf{BA})$, the maximum mutual information $I(\mathbf{x}^\mathrm{DD} ; \mathbf{y}^\mathrm{DD})$, which is also referred to as capacity $C^\mathrm{SISO}$, can be rewritten in the form of (\ref{infor}) as
				\begin{align} \label{infor3}
				C^\mathrm{SISO} & =
				\log _{2}\operatorname{det}\left(\mathbf{I}_{MN}+\frac{\sigma_x^2}{N_{0}} \mathbf{P}^H{\mathbf{H}^\mathrm{DD}}^H (\mathbf{F}_N\otimes\mathbf{F}_M^H){\mathbf{G}^{-\frac{1}{2}}}^H\right.\notag \\&\quad\quad\quad\left.{\mathbf{G}^{-\frac{1}{2}}}(\mathbf{F}_N^H\otimes\mathbf{F}_M)\mathbf{H}^\mathrm{DD}\mathbf{P}\right)\notag\\
				&=\log _{2}\operatorname{det}\left(\mathbf{I}_{MN}+\frac{\sigma_x^2}{N_{0}} \mathbf{P}^H\mathbf{D}^H\mathbf{D}\mathbf{P}\right),
				\end{align}
				where $\mathbf{G}$ is a Hermitian matrix and $\mathbf{D}={\mathbf{G}^{-\frac{1}{2}}}(\mathbf{F}_N^H\otimes\mathbf{F}_M)\mathbf{H}^\mathrm{DD}$.
				Under the proposed EVD-precoded MC-FTN-OTFS signaling scheme, we firstly perform EVD on the matrix $\mathbf{D}^H\mathbf{D}$, which is given by
				\begin{align} \label{lambda}
				\mathbf{D}^H\mathbf{D} = \mathbf{U}_D\mathbf{\Lambda}_D\mathbf{U}_D^H,
				\end{align}
				where  $\mathbf{U}_D\in \mathbb{C}^{MN\times MN}$ is a unitary matrix and $\mathbf{\Lambda}_D\in \mathbb{R}^{MN\times MN}$ is a diagonal matrix with diagonal entries $ [\lambda_{D,0},...,\lambda_{D,MN-1}]$ arranged in descending order.
				Based on the SVD, the precoding matrix $\mathbf{P}$ can be decomposed as:
				\begin{align}
				\mathbf{P} = \mathbf{U}_P{\mathbf{\Lambda}_P}^{\frac{1}{2}}\mathbf{V}_P^H,
				\end{align}
				where the unitary matrices $\mathbf{U}_P\in \mathbb{C}^{MN\times MN}$, $\mathbf{V}_P\in \mathbb{C}^{MN\times MN}$ and the   diagonal matrix $\mathbf{\Lambda}_P\in \mathbb{R}^{MN\times MN}$ are defined similarly to $\mathbf{U}_D$ and $\mathbf{\Lambda}_D$, respectively. 
				Therefore, the capacity is calculated as
				\begin{align}\label{Her}
				C^\mathrm{SISO}\!\! =\!\! \log _{2}\!\operatorname{det} \!\left( \! \mathbf{I}_{MN}\!\!+\!\! \frac{\sigma_x^2}{N_{0}}{\mathbf{\Lambda}_D}^\frac{1}{2}\mathbf{U}_D^H\mathbf{U}_P\mathbf{\Lambda}_P^{\frac{1}{2}} \mathbf{\Lambda}_P^{\frac{1}{2}}\mathbf{U}_P^H  \mathbf{U}_D{\mathbf{\Lambda}_D}^\frac{1}{2}\!\right)\!\! .
				\end{align}
				
				%It can be noted that the matrix ${\mathbf{\Lambda}_D}^\frac{1}{2}\mathbf{U}_D^H\mathbf{U}_P\mathbf{\Lambda}_P \mathbf{\Lambda}_P\mathbf{U}_P^H  \mathbf{U}_D{\mathbf{\Lambda}_D}^\frac{1}{2}$ in (\ref{Her}) can be rewritten as \\ $\left(\mathbf{\Lambda}_P\mathbf{U}_P^H  \mathbf{U}_D{\mathbf{\Lambda}_D}^\frac{1}{2}\right)^H\mathbf{\Lambda}_P\mathbf{U}_P^H  \mathbf{U}_D{\mathbf{\Lambda}_D}^\frac{1}{2}$, which is a positive semi-definite Hermitian matrix.
				
				Based on the Hadamard's inequality $|\mathbf{K}| \leq \prod_{i=1}
				^{MN} \mathbf{K}_{i i}$~\cite{information,information2}, where $\mathbf{K}_{i i}$ denotes the $i$th diagonal element of $\mathbf{K}$. The equality holds if and only if $\mathbf{K}$ is diagonal. Hence, the capacity $C^\mathrm{SISO}$ is realized when  ${\mathbf{\Lambda}_D}^\frac{1}{2}\mathbf{U}_D^H\mathbf{U}_P\mathbf{\Lambda}_P \mathbf{\Lambda}_P\mathbf{U}_P^H  \mathbf{U}_D{\mathbf{\Lambda}_D}^\frac{1}{2}$ is diagonal, which requires that the matrix $\mathbf{U}_D^H\mathbf{U}_P$ is diagonal\cite{EVD}~\cite{Precoding}. 
				%Since $\mathbf{U}_D^H$ and $\mathbf{U}_P$ are both unitary matrixes, matrix $\mathbf{U}_D^H\mathbf{U}_P$ can be diagonal if and only if $\mathbf{U}_D^H\mathbf{U}_P=\mathbf{I}$, where we obtain $\mathbf{U}_P = \mathbf{U}_D$. Therefore, 
				Considering the fact that $\mathbf{U}_D^H$ and $\mathbf{U}_P$ are both unitary matrices, we have the precoding matrix $\mathbf{P}$ as
				\begin{align} \label{Pre}
				\mathbf{P} = \mathbf{U}_D {\mathbf{\Lambda}_P}^{\frac{1}{2}},
				\end{align}
				where ${\mathbf{\Lambda}_P}$ is the PA matrix with diagonal entries $[{\lambda}_{P,0},...,{\lambda}_{P,MN-1}]$. Then, the capacity of our proposed MC-FTN-OTFS signaling for the SISO system is given by
				\begin{align} \label{rate}
				C^\mathrm{SISO} &= \log _{2}\operatorname{det}\left(\mathbf{I}_{MN}+ \frac{\sigma_x^2}{N_{0}}\mathbf{\Lambda}_D^{\frac{1}{2}}{\mathbf{\Lambda}_P} \mathbf{\Lambda}_D^{\frac{1}{2}}\right)  \notag \\
				&=\sum_{k=0}^{MN-1} \log _2\left(1+\frac{\sigma_{{x}}^2}{N_0} \lambda_{P,k} \lambda_{D,k}\right).
				\end{align}

				In the conventional EVD precoding method without PA, ${\mathbf{\Lambda}_P}$ is assumed to be an identity matrix, which means $\mathbf{P} = \mathbf{U}_D$. This method can eliminate the MC-FTN-induced ISI and ICI, but the capacity is not maximized. In the sequel, we propose an optimal PA strategy.

				\subsection{PA Optimization for Capacity Maximization}
				In this subsection, we aim to derive the optimal PA coefficients ${\lambda}_{P,0},...,{\lambda}_{P,MN-1}$ for maximizing  capacity in (\ref{rate}).
				
				Specifically, based on (\ref{rate}), we can formulate the PA problem as
				\begin{align}\label{obj4}
				\begin{aligned}
				\max _{\boldsymbol{\Lambda}_P} \sum_{k=0}^{MN-1} \log _2\left(1+\frac{\sigma_{{x}}^2}{N_0} {\lambda}_{P,k} \lambda_{D,k}\right) ,
				\end{aligned}
				\end{align}
				where $\boldsymbol{\Lambda}_P$ is the PA matrix to be optimized.
				
				{The average transmit energy for an OTFS frame in the  EVD-precoded MC-FTN-OTFS signaling scheme is then given by}
				\begin{align} \label{EN}
				\begin{aligned}
				E_N &=\mathbb{E}\left[\int_{-\infty}^{\infty}|s(t)|^2 d t\right] \\
%				&={{\alpha \beta E_0}}\mathbb{E}\left[\sum_{m_1,n_1} \sum_{m_2,n_2} x_{n_1,m_2} x_{n_2,m_2}^* \right. \\& \left. A_{g_{tx},g_{tx}}(((m_1-m_2)\beta\Delta f),(n_1-n_2)\alpha T)e^{j2\pi m_2\beta\Delta f (n_1-n_2)\alpha T}\right] \\
				&={{\alpha \beta E_0}}\mathbb{E}\left[{\mathbf{x}_P^\mathrm{DD}}^H \mathbf{G} {\mathbf{x}_P^\mathrm{DD}}\right] \\
				%&={{\alpha\beta E_0}}\mathbb{E}\left[\operatorname{tr}\{{{\mathbf{x}_P^\mathrm{DD}}^H \mathbf{G} {\mathbf{x}_P^\mathrm{DD}}}\}\right] \\
				&\overset{(a)}{=}{{\alpha\beta E_0}}\mathbb{E}\left[\operatorname{tr}\{{\mathbf{G}\mathbf{P}{\mathbf{x}^\mathrm{DD}} {\mathbf{x}^\mathrm{DD}}^H}\mathbf{P}^H\}\right]\\
				&={{\alpha\beta E_0}}\operatorname{tr}\{{\mathbf{G}\mathbf{P}\mathbb{E}[{\mathbf{x}^\mathrm{DD}} {\mathbf{x}^\mathrm{DD}}^H}]\mathbf{P}^H\} \\
				&\overset{(b)}{=}{{\alpha\beta E_0\sigma_x^2}}\operatorname{tr}\{\mathbf{G}\mathbf{P}\mathbf{P}^H\},
				\end{aligned}
				\end{align}
				where $(a)$ is obtained due to the fact that $\operatorname{tr}\{\mathbf{AB}\}=\operatorname{tr}\{\mathbf{BA}\}$ and $(b)$ is obtained by $\mathbb{E}\left[{\mathbf{x}^\mathrm{DD}} {\mathbf{x}^\mathrm{DD}}^H\right] = \sigma_{x}^2\mathbf{I}_{MN}$.
				
				We further substitute $\mathbf{P}$ in (\ref{Pre}) into power constraint in (\ref{EN}) and then have
				\begin{align}\label{EN2}
				\begin{aligned}
				E_N &=\alpha\beta E_0 \sigma_{x}^2 \cdot \operatorname{tr}\left\{\mathbf{G} \mathbf{U}_D \mathbf{\Lambda}_{P} \mathbf{ U}_D^H\right\} \\
				&=\alpha\beta E_0 \sigma_{{x}}^2 \cdot \operatorname{tr}\left\{\mathbf{\Lambda}_{P}\underbrace{\mathbf{U}_D^H \mathbf{G} \mathbf{U}_D}_{\boldsymbol{\Phi}}\right\} \\
				&=\alpha\beta E_0 \sigma_{{x}}^2 \sum_{k=0}^{MN-1} \lambda_{P,k} \phi_k,
				\end{aligned}
				\end{align}
				where $\boldsymbol{\Phi}=\mathbf{U}_D^H \mathbf{G U}_D\in \mathbb{C}^{MN\times MN}$ and $\phi_k$ is the $(k+1)$-th diagonal entry of $\boldsymbol{\Phi}$.
				
				Generally, the presence or absence of the precoding matrix should not affect the total transmit energy. Note that the unprecoded MC-FTN-OTFS signaling can be regarded as a special case of the precoded MC-FTN-OTFS scheme, where $\mathbf{P}=\mathbf{I}_{MN}$. In this case,  the transmit energy is calculated as  $E_N={{\alpha }}\beta E_0 MN\sigma_x^2$~\cite{EVD}. In order to maintain the total transmit energy, the precoding matrix should be designed under the energy constraint of $E_N\leq{{\alpha\beta E_0}}MN\sigma_x^2$, which can be further simplified from (\ref{EN2}) as
				\begin{align} \label{cons}
				\sum_{k=0}^{MN-1} {\lambda}_{P,k} \phi_k \leq MN
				\end{align}
				
				%Then we utilize the Lagrange multiplier method to obtain the optimal PA matrix $\mathbf{\Lambda}_P$ that maximizes the mutual information. 
				According to (\ref{obj4}) and (\ref{cons}), the PA problem can be finally formulated as
				\begin{align}
				\begin{aligned}
				&\max _{\boldsymbol{\Lambda}_P} \sum_{k=0}^{MN-1} \log _2\left(1+\frac{\sigma_{{x}}^2}{N_0} {\lambda}_{P,k} \lambda_{D,k}\right) \\
				&\text { s.t. } \sum_{k=0}^{MN-1} {\lambda}_{P,k} \phi_k \leq MN
				\end{aligned}
				\end{align}
				
				We then utilize the Lagrange multiplier method to obtain the optimal PA matrix $\mathbf{\Lambda}_P$ that maximizes the capacity. 
				The Lagrange function is denoted as
				\begin{align}
				J&=\sum_{k=0}^{MN-1} \log _2\left(1+\frac{\sigma_{{x}}^2}{N_0}  {\lambda}_{P,k} \lambda_{D,k}\right) \notag \\ &\quad\quad \quad -\xi\left(\sum_{k=0}^{MN-1}  {\lambda}_{P,k} \phi_k-MN\right),
				\end{align}
				where $\xi$ is the Lagrange multiplier. By taking the first-order derivative of Lagrange function $J$, we can obtain the optimal PA coefficients $ {\lambda}_{P,k}$ as~\cite{EVD}
				%{\color{red}[EVD2,21]}
				\begin{align} \label{gamma}
				{\lambda}_{P,k}=\max \left(\frac{1}{\xi \phi_k \ln 2}-\frac{N_0}{\lambda_{D,k} \sigma_{{x}}^2}, 0\right).
				\end{align}
				Hence, the capacity of our proposed EVD-precoded MC-FTN-OTFS signaling scheme normalized by  {the duration and bandwidth~\cite{Cap1,Cap2}} is denoted as
				\begin{align} \label{C1}
				C^{\mathrm{SISO}}_\mathrm{nor}\!=\!\frac{1}{\alpha \beta M N E_0 }\! \sum_{k=0}^{MN-1}\! \log _2\left(1+\frac{\sigma_{{x}}^2}{N_0} {\lambda}_{P,k} \lambda_{D,k}\right)  \text{[bps/Hz]},
				%\quad[\mathrm{bits} / \mathrm{sec}].
				\end{align}
				where ${\lambda}_{P,k}$ is given by (\ref{gamma}).

		\section{SIC-based MIMO MC-FTN-OTFS Systems}\label{S4}
		For MIMO systems, directly extending the techniques for SISO systems in Section \ref{S3} will lead to a very high computational complexity. In this section, we propose a low-complexity SIC-based precoding scheme for MIMO systems. Specifically, we decompose the capacity maximization problem into $N_T$ subproblems in terms of each data stream. Then we solve the $N_T$ subproblems sequentially and obtain $N_T$ different precoding matrices.

					\subsection{Proposed SIC-based Precoding Scheme for MIMO MC-FTN-OTFS Systems}
					
					Similar to Section \ref{S3}, the mutual information between the received signal $\mathbf{y}^\mathrm{DD}_\mathrm{MIMO}$ and the transmit signal $\mathbf{x}^\mathrm{DD}_\mathrm{MIMO}$ is upper-bounded by
					\begin{align}
					I(\mathbf{x}_\mathrm{P,MIMO}^\mathrm{DD} ; \mathbf{y}_\mathrm{MIMO}^\mathrm{DD}) \leq \log _2 \frac{\operatorname{det}\left({\mathbb{E}\left[\mathbf{y}_\mathrm{MIMO}^\mathrm{DD}{\mathbf{y}_\mathrm{MIMO}^\mathrm{DD}}^H\right]}\right)}{\operatorname{det}\left(\mathbb{E}\left[\mathbf{z}_\mathrm{MIMO}^\mathrm{DD}{\mathbf{z}_\mathrm{MIMO}^\mathrm{DD}}^H\right]\right)}.
					\end{align}
					
					The correlation matrix of $\mathbf{y}_\mathrm{MIMO}^\mathrm{DD}$ is calculated as (\ref{corr41}) at the top of the next page,
					\begin{figure*}[!h]
						\centering
					\begin{align}\label{corr41}
					&\mathbb{E}\left[\mathbf{y}_\mathrm{MIMO}^\mathrm{DD}{\mathbf{y}_\mathrm{MIMO}^\mathrm{DD}}^H\right] \notag \\ &=\mathbb{E}\left[(\mathbf{H}^\mathrm{DD}_\mathrm{MIMO} \mathbf{P}_\mathrm{MIMO}\mathbf{x}^\mathrm{DD}_\mathrm{MIMO} +\mathbf{z}_\mathrm{MIMO}^\mathrm{DD})(\mathbf{H}^\mathrm{DD}_\mathrm{MIMO} \mathbf{P}_\mathrm{MIMO}\mathbf{x}^\mathrm{DD}_\mathrm{MIMO} +\mathbf{z}_\mathrm{MIMO}^\mathrm{DD})^H\right]\notag \\ 
					&=\mathbf{H}^\mathrm{DD}_\mathrm{MIMO}\mathbf{P}_\mathrm{MIMO}\mathbb{E}\left[\mathbf{x}^\mathrm{DD}_\mathrm{MIMO}{\mathbf{x}^\mathrm{DD}_\mathrm{MIMO}}^H\right]\mathbf{P}_\mathrm{MIMO}^H{\mathbf{H}^\mathrm{DD}_\mathrm{MIMO}}^H +\mathbb{E}\left[\mathbf{z}_\mathrm{MIMO}^\mathrm{DD}{\mathbf{z}_\mathrm{MIMO}^\mathrm{DD}}^H\right],
					\end{align} 
					\vspace{-6mm}
					\hrulefill
					\end{figure*}
					where $\mathbb{E}\left[\mathbf{z}_\mathrm{MIMO}^\mathrm{DD}{\mathbf{z}_\mathrm{MIMO}^\mathrm{DD}}^H\right]$ is the correlation matrix of $\mathbf{z}_\mathrm{MIMO}^\mathrm{DD}$.
					In order to calculate $\mathbb{E}\left[\mathbf{z}_\mathrm{MIMO}^\mathrm{DD}{\mathbf{z}_\mathrm{MIMO}^\mathrm{DD}}^H\right]$, we firstly assume that $\mathbf{z}_{n_r}^\mathrm{DD}$s are independent among $N_r$ receive antennas, then we have the correlation matrix of $\mathbf{z}^\mathrm{DD}_{n_r}$ according to (\ref{zcorr}) as
					\begin{align}\label{zcorr2}
					\mathbb{E}\left[\mathbf{z}_{n_r}^\mathrm{DD}{\mathbf{z}_{n_r}^\mathrm{DD}}^H\right] = N_0(\mathbf{F}_N\otimes\mathbf{F}_M^H)\mathbf{G}(\mathbf{F}_N^H\otimes\mathbf{F}_M),
					\end{align}
					where $\mathbf{G}$ is the MC-FTN-induced ISI and ICI matrix.
					Hence, the correlation matrix $\mathbb{E}\left[\mathbf{z}_\mathrm{MIMO}^\mathrm{DD}{\mathbf{z}_\mathrm{MIMO}^\mathrm{DD}}^H\right]$ is given by 
					\begin{align}
					\mathbb{E}\left[\mathbf{z}_\mathrm{MIMO}^\mathrm{DD}{\mathbf{z}_\mathrm{MIMO}^\mathrm{DD}}^H\right] = N_0 (\mathbf{I}_{N_R}\otimes \mathbf{F}_N\otimes\mathbf{F}_M^H)(\mathbf{I}_{N_R}\otimes\mathbf{G})\notag\\(\mathbf{I}_{N_R}\otimes \mathbf{F}_N^H\otimes\mathbf{F}_M),
					\end{align}
					
					Then we have the following correlation matrix of $\mathbf{y}_\mathrm{MIMO}^\mathrm{DD}$
					\begin{align}
					\mathbb{E}\left[\mathbf{y}_\mathrm{MIMO}^\mathrm{DD}{\mathbf{y}_\mathrm{MIMO}^\mathrm{DD}}^H\right]  
					=
					\sigma_{x}^2 \mathbf{H}^\mathrm{DD}_\mathrm{MIMO} \mathbf{P}_\mathrm{MIMO}{\mathbf{P}_\mathrm{MIMO}}^H{\mathbf{H}^\mathrm{DD}_\mathrm{MIMO}}^H \notag \\ +N_0 (\mathbf{I}_{N_R}\otimes \mathbf{F}_N\otimes\mathbf{F}_M^H)(\mathbf{I}_{N_R}\otimes\mathbf{G})(\mathbf{I}_{N_R}\otimes \mathbf{F}_N^H\otimes\mathbf{F}_M),
					\end{align} 
					where  $\mathbb{E}\left[\mathbf{x}^\mathrm{DD}_\mathrm{MIMO}{\mathbf{x}^\mathrm{DD}_\mathrm{MIMO}}^H\right]=\sigma_x^2\mathbf{I}_{MNN_T}$.

					Hence, the maximum mutual information, which is referred to as the capacity $C^\mathrm{MIMO}$, is represented as (\ref{inforMIMO}), which is shown at the top of the next page.
					\begin{figure*}
						\centering
%						\vspace{-3mm}
					\begin{align} \label{inforMIMO}
					C^\mathrm{MIMO} 
					%&\leq \log _2 \frac{\operatorname{det}\left(\sigma_{x}^2 \mathbf{H}^\mathrm{DD}_\mathrm{MIMO} \mathbf{P}_\mathrm{MIMO}{\mathbf{P}_\mathrm{MIMO}}^H{\mathbf{H}^\mathrm{DD}_\mathrm{MIMO}}^H +N_0 (\mathbf{I}_{N_R}\otimes \mathbf{F}_N\otimes\mathbf{F}_M^H)(\mathbf{I}_{N_R}\otimes\mathbf{G})(\mathbf{I}_{N_R}\otimes \mathbf{F}_N^H\otimes\mathbf{F}_M)\right)}{\operatorname{det}\left(N_0 (\mathbf{I}_{N_R}\otimes \mathbf{F}_N\otimes\mathbf{F}_M^H)(\mathbf{I}_{N_R}\otimes\mathbf{G})(\mathbf{I}_{N_R}\otimes \mathbf{F}_N^H\otimes\mathbf{F}_M)\right)}\notag\\
					&= \log _{2}\operatorname{det}\left(\mathbf{I}_{MNN_R}+\frac{\sigma_x^2}{N_{0}}(\mathbf{I}_{N_R}\otimes\mathbf{G}^{-1})(\mathbf{I}_{N_R}\otimes \mathbf{F}_N^H\otimes\mathbf{F}_M)\mathbf{H}^\mathrm{DD}_\mathrm{MIMO} \mathbf{P}_\mathrm{MIMO}\right. \notag \\
					&\left.\quad \quad \quad {\mathbf{P}_\mathrm{MIMO}}^H{\mathbf{H}^\mathrm{DD}_\mathrm{MIMO}}^H(\mathbf{I}_{N_R}\otimes \mathbf{F}_N\otimes\mathbf{F}_M^H)\right) \notag \\
					&=\log _{2}\operatorname{det}\left(\mathbf{I}_{MNN_R}+\frac{\sigma_x^2}{N_{0}} (\mathbf{I}_{N_R}\otimes \mathbf{G}^{-\frac{1}{2}} )(\mathbf{I}_{N_R}\otimes \mathbf{F}_N^H\otimes\mathbf{F}_M)\mathbf{H}^\mathrm{DD}_\mathrm{MIMO} \mathbf{P}_\mathrm{MIMO} \right.\notag\\&\left. \quad \quad \quad {\mathbf{P}_\mathrm{MIMO}}^H{\mathbf{H}^\mathrm{DD}_\mathrm{MIMO}}^H(\mathbf{I}_{N_R}\otimes \mathbf{F}_N\otimes\mathbf{F}_M^H)(\mathbf{I}_{N_R}\otimes \mathbf{G}^{-\frac{1}{2}} )\right).
					\end{align}
					\vspace{-3mm}
					\hrulefill
					\end{figure*}
					
					By setting $\mathbf{D}_\mathrm{MIMO}=(\mathbf{I}_{N_R}\otimes \mathbf{G}^{-\frac{1}{2}} )(\mathbf{I}_{N_R}\otimes \mathbf{F}_N^H\otimes\mathbf{F}_M)\mathbf{H}^\mathrm{DD}_\mathrm{MIMO}$, the function (\ref{inforMIMO}) can be denoted as
					\begin{align} \label{PMIMO}
					C^\mathrm{MIMO}\!\!\! =\!\!  \log _{2}\!\operatorname{det}\!\!\left(\!\!\mathbf{I}_{MNN_R}\!\!\!+\!\! \frac{\sigma_x^2}{N_{0}}\!\mathbf{D}_\mathrm{MIMO} \mathbf{P}_\mathrm{MIMO}{\mathbf{P}^H_\mathrm{MIMO}}\mathbf{D}_\mathrm{MIMO}^H\!\right)\!\!,
					\end{align}
					where $\mathbf{P}_\mathrm{MIMO}=\operatorname{diag}\{\mathbf{P}_{1},\ldots, \mathbf{P}_{N_T} \}$ is a block diagonal matrix. 
					Then we have the capacity maximization problem for the MIMO system		
					\begin{align} \label{optMIMO}
					\begin{aligned}
					\max _{{\mathbf{P}}_\mathrm{MIMO}}   \log _{2}\!\operatorname{det}\!\left(\!\mathbf{I}_{MNN_R}\!\!\!+\! \frac{\sigma_x^2}{N_{0}}\!\mathbf{D}_\mathrm{MIMO} \mathbf{P}_\mathrm{MIMO}{\mathbf{P}^H_\mathrm{MIMO}}\mathbf{D}_\mathrm{MIMO}^H\right) \!
					\end{aligned}.
					\end{align}
					
						\begin{figure}[!t]
						\center{\includegraphics[width=1\columnwidth]{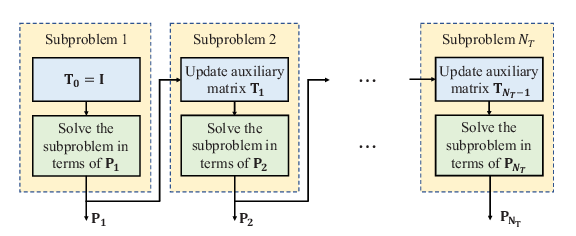}}
						\vspace{-3mm}
						\caption{{The diagram of the proposed SIC-based precoding scheme.}}
						\label{diagram3}
						\vspace{-3mm}
					\end{figure}
					
					{\color{red}Recall Section III, if we adopt the EVD precoding method to solve (\ref{optMIMO}), 
					the computational complexity  is in the order of $\mathcal{O}\left((MNN_T)^3\right)$, which is extremely excessive. Note that  the EVD precoding is referred to as the WF-based precoding after power allocation in Section {\ref{S5}}.
				 Motivated by the SIC method, we will divide the $MNN_T\times MNN_T$ block diagonal matrix $\mathbf{P}_\mathrm{MIMO}$ into $N_T$ matrices of dimension $MN\times MN$ and decompose the problem (\ref{optMIMO}) with the complexity of $\mathcal{O}\left((MNN_T)^3\right)$ into $N_T$ subproblems, each of which can be solved with the complexity of $\mathcal{O}\left((MN)^3\right)$.}

				{\color{red}To be specific, we divide the block diagonal precoding matrix $\mathbf{P}_\mathrm{MIMO}$ into $\mathbf{P}_\mathrm{MIMO} = \left[\widetilde{\mathbf{P}}_{N_T-1}\ \overline{\mathbf{P}}_{N_T}\right]$, where $\widetilde{\mathbf{P}}_{N_T-1} \in \mathbb{C}^{MNN_T\times MN(N_T-1)}$ is constructed by the first $MN(N_T-1)$ columns of $\mathbf{P}_\mathrm{MIMO}$ and  $\overline{\mathbf{P}}_{N_T} \in \mathbb{C}^{MNN_T\times MN}$ is constructed by  the last $MN$ columns of ${\mathbf{P}}_\mathrm{MIMO}$.}  Hence, the capacity of the MIMO MC-FTN-OTFS system is calculated as (\ref{capMIMO}) at the top of the next page, 
				\begin{figure*}[!h]
					\centering
%					\vspace{-5mm}
					\begin{align} \label{capMIMO}
					C^\mathrm{MIMO} \notag &=
					\log _{2}\operatorname{det}\left(\mathbf{I}_{MNN_T}+ \frac{\sigma_x^2}{N_{0}}\mathbf{D}_\mathrm{MIMO} \mathbf{P}_\mathrm{MIMO}\mathbf{P}_\mathrm{MIMO}^H\mathbf{D}_\mathrm{MIMO}^H\right) \notag \\
					&=\log _{2}\operatorname{det}\left(\mathbf{I}_{MNN_R}+ \frac{\sigma_x^2}{N_{0}}\mathbf{D}_\mathrm{MIMO} \left[\widetilde{\mathbf{P}}_{N_T-1}\ \overline{\mathbf{P}}_{N_T}\right]\left[\widetilde{\mathbf{P}}_{N_T-1}\ \overline{\mathbf{P}}_{N_T}\right]^H\mathbf{D}_\mathrm{MIMO}^H\right)\notag \\
					&=\log _{2}\operatorname{det}\left(\mathbf{I}_{MNN_R}+ \frac{\sigma_x^2}{N_{0}}\mathbf{D}_\mathrm{MIMO} \widetilde{\mathbf{P}}_{N_T-1}\widetilde{\mathbf{P}}_{N_T-1}^H\mathbf{D}_\mathrm{MIMO}^H+\frac{\sigma_x^2}{N_{0}}\mathbf{D}_\mathrm{MIMO} \overline{\mathbf{P}}_{N_T}\overline{\mathbf{P}}_{N_T}^H\mathbf{D}_\mathrm{MIMO}^H\right) \notag \\
					&\overset{(a)}{=}\log _{2}\operatorname{det}\left(\mathbf{T}_{N_T-1}\right)+\log _{2}\operatorname{det}\left(\mathbf{I}_{MNN_R}+\frac{\sigma_x^2}{N_{0}}\mathbf{T}_{N_T-1}^{-1}\mathbf{D}_\mathrm{MIMO} \overline{\mathbf{P}}_{N_T}\ \overline{\mathbf{P}}_{N_T}^H\mathbf{D}_\mathrm{MIMO}^H\right) \notag \\
					&\overset{(b)}{=}\log _{2}\operatorname{det}\left(\mathbf{T}_{N_T-1}\right)+\log _{2}\operatorname{det}\left(\mathbf{I}_{MN}+\frac{\sigma_x^2}{N_{0}}\overline{\mathbf{P}}_{N_T}^H\mathbf{D}_\mathrm{MIMO}^H\mathbf{T}_{N_T-1}^{-1}\mathbf{D}_\mathrm{MIMO} \overline{\mathbf{P}}_{N_T}\ \right),
					\end{align}
					\hrulefill
					\end{figure*}
					where $(a)$ is calculated by defining  $\mathbf{T}_{N_T-1}=\mathbf{I}_{MNN_R}+ \frac{\sigma_x^2}{N_{0}}\mathbf{D}_\mathrm{MIMO} \widetilde{\mathbf{P}}_{N_T-1}\widetilde{\mathbf{P}}_{N_T-1}^H\mathbf{D}_\mathrm{MIMO}^H$ and $(b)$ is obtained due to $\operatorname{det}\left(\mathbf{I}+\mathbf{AB}\right)=\operatorname{det}\left(\mathbf{I}+\mathbf{BA}\right)$. Similarly, we can further decompose $\log _{2}\operatorname{det}\left(\mathbf{T}_{N_T-1}\right)$  as (\ref{SIC}) at the top of the next page, 
					\begin{figure*}[!h]
						\centering
						\vspace{-5mm}
					\begin{align}\label{SIC}
					\log _{2}\operatorname{det}\left(\mathbf{T}_{N_T-1}\right) =\log _{2}\operatorname{det}\left(\mathbf{T}_{N_T-2}\right)+\log _{2}\operatorname{det}\left(\mathbf{I}_{MN}+\frac{\sigma_x^2}{N_{0}}\overline{\mathbf{P}}_{N_T-1}^H\mathbf{D}_\mathrm{MIMO}^H\mathbf{T}_{N_T-2}^{-1}\mathbf{D}_\mathrm{MIMO} \overline{\mathbf{P}}_{N_T-1}\right),
					\end{align}
					\hrulefill
%					\vspace{-5mm}
					\end{figure*}
					where $\overline{\mathbf{P}}_{N_T-1}\in \mathbb{C}^{MNN_T\times MN}$ is constructed by the last $MN$ columns of $\widetilde{\mathbf{P}}_{N_T-1}$. {\color{red} Without loss of generality, we use  $\overline{\mathbf{P}}_{n_t}$ to denote the matrix formed by the last $MN$ columns of $\widetilde{\mathbf{P}}_{n_t}$ and use $\widetilde{\mathbf{P}}_{n_t-1}$ to denote the matrix constructed by the remaining columns of $\widetilde{\mathbf{P}}_{n_t}$ after separating the last $MN$ columns.}
					Therefore, after $N_T$ such decompositions, the capacity of this MIMO MC-FTN-OTFS system can be expressed as
					\begin{align} \label{submutual}
					C^\mathrm{MIMO}\!\!\! = \!\!\!
					\sum_{n_t=1}^{N_T}\!\! \log _{2} \!\operatorname{det}\!\left(\!\mathbf{I}_{MN}\!\!+\!\!\frac{\sigma_x^2}{N_{0}}\!\overline{\mathbf{P}}_{\!n_t}^H\mathbf{D}_\mathrm{MIMO}^H\mathbf{T}_{n_t\!-\!1}^{-1} \mathbf{D}_\mathrm{MIMO}\overline{\mathbf{P}}_{n_t}\!\right)\!\!,
					\end{align}
					where $\overline{\mathbf{P}}_{n_t} = \left[\mathbf{0}_{(n_t-1)MN\times MN}^T, {{\mathbf{P}}_{n_t}}^T,\mathbf{0}_{(N_T-n_t)MN\times MN}^T \right]^T$, $\mathbf{T}_{n_t-1}=\mathbf{I}_{MNN_R}+ \frac{\sigma_x^2}{N_{0}}\mathbf{D}_\mathrm{MIMO} \overline{\mathbf{P}}_{n_t-1}\overline{\mathbf{P}}_{n_t-1}^H$ $\mathbf{D}_\mathrm{MIMO}^H$ and $\mathbf{T}_0=\mathbf{I}$. {\color{red} Motivated by the idea of the SIC signal detection, we intend to decompose the capacity maximization problem in (\ref{submutual}) into $N_T$ subproblems, each corresponding to a DD domain data stream. Specifically, Fig. \ref{diagram3} illustrates the workflow of the proposed SIC-based precoding scheme. Initially, we initialize the auxiliary matrix $\mathbf{T}_0$ and then solve the subproblem corresponding to the  first data stream to obtain the precoding matrix $\mathbf{P}_1$. Next, the solution from solving the first subproblem is used to update the auxiliary matrix $\mathbf{T}_1$, which allows us to solve the next subproblem and obtain $\mathbf{P}_2$. After repeating this step $N_T$ times, we can obtain all $N_T$ precoding matrices.}
					
					\subsection{Optimal Solution to the Sub-Capacity Maximization Problem}
					
					In this subsection, we show how to solve $N_T$ optimization problems. Assuming $\overline{\mathbf{Q}}_{n_t}=\mathbf{D}_\mathrm{MIMO}^H\mathbf{T}_{n_t}^{-1}\mathbf{D}_\mathrm{MIMO}$, the capacity can be written as
					\begin{align} \label{sub1}
					C^\mathrm{MIMO} =
					\sum_{n_t=1}^{N_T} \log _{2}\operatorname{det}\left(\mathbf{I}_{MN}+\frac{\sigma_x^2}{N_{0}} \overline{\mathbf{P}}_{n_t}^H\overline{\mathbf{Q}}_{n_t-1}\overline{\mathbf{P}}_{n_t}\right),
					\end{align}
					where only the columns from $(n_t-1)MN+1$ to $n_tMN$ of  $\overline{\mathbf{P}}_{n_t}$ are non-zero. 
					Let $P_{n_t}$, $n_t=1,\ldots,N_T$ denote the power constraint {\color{red}corresponding} to the $n_t$-th data stream. 
					Hence, the transmit energy of the $n_t$-th data stream is represented as
					\begin{align} 
					\begin{aligned}
					E_N &={{\alpha \beta E_0}}\mathbb{E}\left[\mathbf{x}_{\mathrm{P},n_t}^H \mathbf{G} \mathbf{x}_{\mathrm{P},n_t}\right] \\
					&={{\alpha \beta E_0}}\sigma_x^2\operatorname{tr}\left(\mathbf{P}_{n_t} {\mathbf{P}_{n_t}}^H \mathbf{G }\right), n_t=1,\ldots,N_T.
					% \leq P_{n_t} .
					\end{aligned}
					\end{align}
					Therefore, the SIC-based precoding subproblem corresponding to the $n_t$-th data stream can be formulated as
					\begin{align} \label{opt}
					\begin{aligned}
					&\max _{\overline{\mathbf{P}}_{n_t}} \quad  \overline{F}(\overline{\mathbf{P}}_{n_t}) = \log _{2}\operatorname{det}\left(\mathbf{I}_{MNN_R}+\frac{\sigma_x^2}{N_{0}} \overline{\mathbf{P}}_{n_t}^H\overline{\mathbf{Q}}_{n_t-1}\overline{\mathbf{P}}_{n_t}\right) \\
					&\text { s.t. } \quad \operatorname{tr}\left(\mathbf{P}_{n_t} {\mathbf{P}_{n_t}}^H \mathbf{G }\right) \leq P_{n_t}/(\alpha \beta E_0 \sigma_x^2) 
					\end{aligned},
					\end{align}
					where we use $\overline{F}(\overline{\mathbf{P}}_{n_t})$ to denote the objective function corresponding to $\overline{\mathbf{P}}_{n_t}$. 
					
					According to \cite{SIC1}, the objective function $\overline{F}(\overline{\mathbf{P}}_{n_t})$ in (\ref{opt}) can be simplified as
					\begin{align} \label{obj}
					{F}(\overline{\mathbf{P}}_{n_t}) =  \log _{2}\operatorname{det}\left(\mathbf{I}_{MNN_R}+\frac{\sigma_x^2}{N_{0}} {\mathbf{P}}_{n_t}^H{\mathbf{Q}}_{n_t-1}{\mathbf{P}}_{n_t}\right),
					\end{align}
					where ${F}(\overline{\mathbf{P}}_{n_t})$ denotes the simplified objective function and ${\mathbf{Q}}_{n_t-1} \in \mathbb{C}^{MN\times MN}$ is the submatrix of $\overline{\mathbf{Q}}_{n_t-1}$ consisting of the $((n_t-1)MN+1)$-th to $n_tMN$-th rows and columns, which can be denoted as
					\begin{align} \label{Q}
					{\mathbf{Q}}_{n_t-1} = \mathbf{R}\overline{\mathbf{Q}}_{n_t-1}\mathbf{R}^H,
					\end{align}
					where $\mathbf{R} = \operatorname{diag}\{\mathbf{0}_{(n_t-1)MN\times MN}, \mathbf{I}_{MN},\mathbf{0}_{(N_T-n_t)MN\times MN} \}$. We transform the high-dimensional matrix $\overline{\mathbf{P}}_{n_t}$ in (\ref{opt}) into the $MN\times MN$ precoding matrix $\mathbf{P}_{n_t}$ in (\ref{obj}) corresponding to the $n_t$-th data stream. 
					
					Since $\mathbf{Q}_{n_t-1}$ is Hermitian, it can be decomposed by EVD as $\mathbf{Q}_{n_t-1} = \mathbf{U}_{n_t-1}\mathbf{\Lambda}_{Q,n_t-1}\mathbf{U}_{n_t-1}^H$, where $\mathbf{U}_{n_t-1}$ is a unitary matrix and $\mathbf{\Lambda}_{Q,n_t-1}$ is a diagonal matrix whose diagonal elements $[\mathbf{\lambda}_{Q,n_t-1,1},\ldots,\mathbf{\lambda}_{Q,n_t-1,MN}]$ are arranged in descending order. Similar to the SISO case, the optimal $\mathbf{P}_{n_t}$ is given by $\mathbf{P}_{n_t}=\mathbf{U}_{n_t-1}\mathbf{\Gamma}_{n_t}^\frac{1}{2}$~\cite{Precoding}, where $\mathbf{\Gamma}_{n_t}$ denotes the diagonal PA matrix of the $n_t$-th antenna with diagonal elements $[\mathbf{\gamma}_{n_t,1},\ldots,\mathbf{\gamma}_{n_t,MN}]$. Then the objective function in (\ref{obj}) can be denoted as
					\begin{align} \label{obj3}
					\max _{\mathbf{\gamma}_{n_t,i}} \quad \sum_{i=1}^{MN} \log _{2}(1+\frac{\sigma_x^2}{N_{0}} \mathbf{\gamma}_{n_t,i}\mathbf{\lambda}_{Q,n_t-1,i}), 
					\end{align}

					The total power constraint can be simplified as
					\begin{align} 
					\operatorname{tr}\left(\mathbf{P}_{n_t} {\mathbf{P}_{n_t}}^H \mathbf{G }\right)&=\operatorname{tr}\left(\mathbf{U}_{n_t-1}\mathbf{\Gamma}_{n_t} {\mathbf{U}_{n_t-1}}^H \mathbf{G }\right)\notag \\
					&\overset{(a)}{=}\operatorname{tr}\left(\mathbf{\Gamma}_{n_t} \underbrace{{\mathbf{U}_{n_t-1}}^H \mathbf{G }\mathbf{U}_{n_t-1}}_{\mathbf{\Psi}_{n_t-1}}\right)\notag \\
					&=\sum_{i=1}^{MN}{\gamma_{n_t,i}\psi_{n_t-1,i}} \leq P_{n_t}/(\alpha \beta E_0 \sigma_x^2),
					\end{align}
					where $(a)$ is obtained due to the facts  $\operatorname{tr}(\mathbf{AB})=\operatorname{tr}(\mathbf{BA})$ and $\psi_{n_t-1,i}$ is the $i$-th elements of $ \mathbf{\Psi}_{n_t-1}$.
					Hence, (\ref{opt}) can be written as
					\begin{align} \label{opt2}
					\begin{aligned}
					&\max _{\{{\gamma}_{n_t,i}\}_{n_t=1,\ldots,N_T,i=1,\ldots,MN}} \quad \sum_{i=1}^{MN} \log _{2}(1+\frac{\sigma_x^2}{N_{0}} {\gamma}_{n_t,i}{\lambda}_{Q,n_t-1,i}), \\
					&\text { s.t. } \quad \sum_{i=1}^{MN}{\gamma_{n_t,i}\psi_{n_t-1,i}}\leq P_{n_t}/(\alpha \beta E_0 \sigma_x^2), \ \ n_t=1,\ldots,N_T  . 
					\end{aligned}
					\end{align}
					Similarly, by utilizing the Lagrange multiplier method in Section \ref{S3} from the $1$-st subproblem to the $N_T$-th subproblem, we can obtain the PA coefficients  ${\gamma}_{n_t,i}$, $i=1,\ldots,MN$ for $n_t=1,\ldots,N_T$ as
					\begin{align} \label{Gamma}
					{\gamma}_{n_t,i}=\max \left(\frac{1}{\xi_{n_t} \psi_{n_t-1,i} \ln 2}-\frac{N_0}{{\lambda}_{Q,n_t-1,i} \sigma_{{x}}^2}, 0\right),
					\end{align}
					where $\xi_{n_t}$ is the Lagrange multiplier corresponding to the $n_t$-th subproblem.
					Then the capacity normalized by the duration and bandwidth of the proposed MIMO MC-FTN-OTFS system is given by~\cite{Cap1,Cap2}
					\begin{align} \label{gammaMIMO}
					C^\mathrm{MIMO}_\mathrm{nor}=\frac{1}{\alpha \beta M N E_0 } \! \sum_{n_t=1}^{N_T} \! \sum_{i=1}^{MN} \log _2\left(1+\frac{\sigma_{{x}}^2}{N_0} \gamma_{n_t,i} \lambda_{Q,n_t-1,i}\right).
					%\quad[\mathrm{bps} / \mathrm{Hz}].
					\end{align}
					\vspace{-2mm}
					
					To summarize the proposed SIC-based precoding scheme for the MIMO MC-FTN-OTFS system at a glance, we present it in \textbf{Algorithm 1}, which can be explained as follows. We firstly compute the matrix $\mathbf{Q}_{n_t-1}$ using the initial auxiliary matrix $\mathbf{T}_{n_t-1}$ and then perform EVD on $\mathbf{Q}_{n_t-1}$ to obtain the unitary matrix $\mathbf{U}_{n_t-1}$. After that we solve the sub-capacity maximization problems in (\ref{opt2}) and obtain the PA coefficients $\gamma_{n_t,i}$ as shown in (\ref{Gamma}). The precoding matrix corresponding to the $n_t$-th data stream is represented as $\mathbf{P}_{n_t}=\mathbf{U}_{n_t-1}\mathbf{\Gamma}_{n_t}^{\frac{1}{2}}$. Then we update the auxiliary matrix $\mathbf{T}_{n_t}$ by using $\mathbf{P}_{n_t}$. After repeating this iteration for $N_T$ times, we can obatin $N_t$ precoding matrices $\mathbf{P}_{n_t}$ for $n_t=1,\ldots,N_T$.

					\begin{algorithm}[!t]
						\caption{The proposed SIC-based precoding scheme.}
						
						\KwIn{(1) Initial auxiliary matrix $\mathbf{T}_0=\mathbf{I}$.	
							
							\quad\quad\quad
							(2) The number of subproblems $N_T$.}

							\For{$1\leq n_t\leq N_T$}{\begin{algorithmic}[1]

						\STATE Compute $\overline{\mathbf{Q}}_{n_t-1}=\mathbf{D}_\mathrm{MIMO}^H\mathbf{T}_{n_t-1}^{-1}\mathbf{D}_\mathrm{MIMO}$, ${\mathbf{Q}}_{n_t-1} = \mathbf{R}\overline{\mathbf{Q}}_{n_t-1}\mathbf{R}^H$ in (\ref{Q})
						\STATE Compute $\mathbf{U}_{n_t-1}$ by performing EVD on matrix\\ ${\mathbf{Q}}_{n_t-1}$.
						\STATE Compute $\mathbf{\Gamma}_{n_t}$ by (\ref{Gamma}).
						\STATE $\mathbf{P}_{n_t}=\mathbf{U}_{n_t-1}\mathbf{\Gamma}_{n_t}^{\frac{1}{2}}$.
						\STATE Update auxiliary matrix $\mathbf{T}_{n_t}=\mathbf{I}_{MNN_R}+$ $ \frac{\sigma_x^2}{N_{0}}\mathbf{D}_\mathrm{MIMO} \overline{\mathbf{P}}_{n_t}\overline{\mathbf{P}}_{n_t}^H\mathbf{D}_\mathrm{MIMO}^H$, where $\overline{\mathbf{P}}_{n_t}\!\! =\!\! \left[\mathbf{0}_{(n_t-1)MN\times MN}^T,\! {{\mathbf{P}}_{n_t}}^T,\!\mathbf{0}_{(N_T-n_t)MN\times MN}^T \right]^T\!\!$.

							\end{algorithmic}}

						\KwOut{Precoding matrices $\mathbf{P}_{n_t},\ n_t=1,\ldots,N_T $}
					\end{algorithm}

	\section{Numerical Results} \label{S5}
					\begin{figure*}[t]
						\center{\includegraphics[width=1.8\columnwidth]{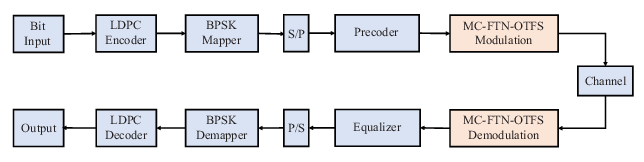}}
						\caption{\color{red}The signal processing flowchart  of the proposed MC-FTN-OTFS transmitter and receiver.}
						\label{diagramLDPC}
						\vspace{-7mm}
					\end{figure*}

					%评估了什么，比较了什么，什么作为了benchmark
					In this section, we present numerical  results of  our proposed MC-FTN-OTFS precoding scheme for SISO and MIMO systems to evaluate the performance. In the following simulation, the parameters are described as follows. The Nyquist time interval is set as $T_0 = 1$ and the transmit power per subcarrier is set as $E_0=1$. The channel is generated in the DD domain with randomly generated channel coefficients according to the complex Gaussian distribution $\mathcal{CN}(0,1/L)$, where we assume $L=3$. For the sake of simplicity, we set the dimension of the data block as $8\times 4$. If not otherwise specified, the roll-off factor of the RRC shaping filter is $\theta = 0.25$. The achievable information rate is the average of 10000 random realizations.
					
					{\color{red} Fig. {\ref{diagramLDPC}} illustrates the signal processing flowchart  of the proposed MC-FTN-OTFS transmitter and receiver. At the transmitter, a low-density parity-check (LDPC)  encoder is employed to ensure high-reliability information transmission. The encoded bits are then mapped using binary phase shift keying (BPSK) and divided into $M$ groups corresponding to $M$ subcarriers.  After passing through the serial-to-parallel (S/P) module and the precoder, the signals undergo MC-FTN-OTFS modulation.
						At the receiver, the received signal firstly undergoes the MC-FTN-OTFS demodulation, followed by the equalization. After passing through the parallel-to-serial (P/S) module and the BPSK demapper, the signals are finally decoded by the LDPC decoder.}

					We also compare the performance of our proposed scheme with existing schemes. We compare our proposed MC-FTN-OTFS signaling scheme with the following benchmarks: 1) \textbf{\textit{Conventional FTN signaling}}~\cite{EVD,PRE3}: The non-orthogonal pulses are only adopted in time domain; 
					%2) \textbf{Ultra-dense NOFDM signaling}: The non-orthogonal pulses are only adopted in frequency domain. 
					2) \textbf{\textit{Classical Nyquist-criterion-based scheme}}~\cite{OTFS1}: The orthogonal pulses are used in both time and frequency domains;
					 3) \textbf{\textit{Classical water-filling (WF)-based  precoding scheme for MIMO systems}}~\cite{WF}: The WF-based precoding is directly performed on the MIMO capacity maximization problem (\ref{optMIMO}}), which achieves the optimal capacity performance.
				%检查一下
					Note that benchmarks 1) and 2) are adopted in both SISO and MIMO systems, and benchmark 3) is only considered in the MIMO system.

					\subsection {MC-FTN-OTFS Signaling Scheme for the SISO System}
					In this subsection, we mainly focus on the performance of MC-FTN-OTFS signaling for the SISO system. Simulation results of  capacity and BER of our proposed scheme are presented.  
					
					%\begin{figure}[t]
					%	\center{\includegraphics[width=0.6\columnwidth]{Ralpha.eps}}
					%	\caption{The normalized information rate of our proposed FTN signaling for OTFS scheme. The roll-off factor of the RRC shaping filter if $\theta = 0.25$. The compression factor is $\alpha = 1,0.9,0.8$. The block size of OTFS frame in the DD domain is $8\times4$.}
					%	\label{Ralpha}
					%\end{figure}

					%Fig. \ref{Ralpha} shows the normalized information rate for the special case of FTN signaling for OTFS systems with compression factors as $\alpha$ $=$ $1$, $0.9$ and $0.8=1/(1+\theta)$ with and without power allocation (PA). The precoding matrix with power allocation is $\mathbf{P} = \mathbf{U}_{D} \boldsymbol{\Gamma}^{\frac{1}{2}}$, while the precoding matrix without power allocation is $\mathbf{P} = \mathbf{U}_{D}$, which is known as the conventional SVD-based precoding.
					% The data block size $M\times N$ is set to $8\times4$ and the number of paths is set to $P=3$. In Fig. \ref{Ralpha}, we observe that the normalized information rate of FTN signaling for OTFS systems with PA is higher than the counterparts without PA. Furthermore, the information rate with FTN signaling is higher than that of the Nyquist-based OTFS scheme~\cite{ConSVD} and the information rate increases as the compression factor $\alpha$ decreases.
					
				\begin{figure}[t]
						\center{\includegraphics[width=0.8\columnwidth]{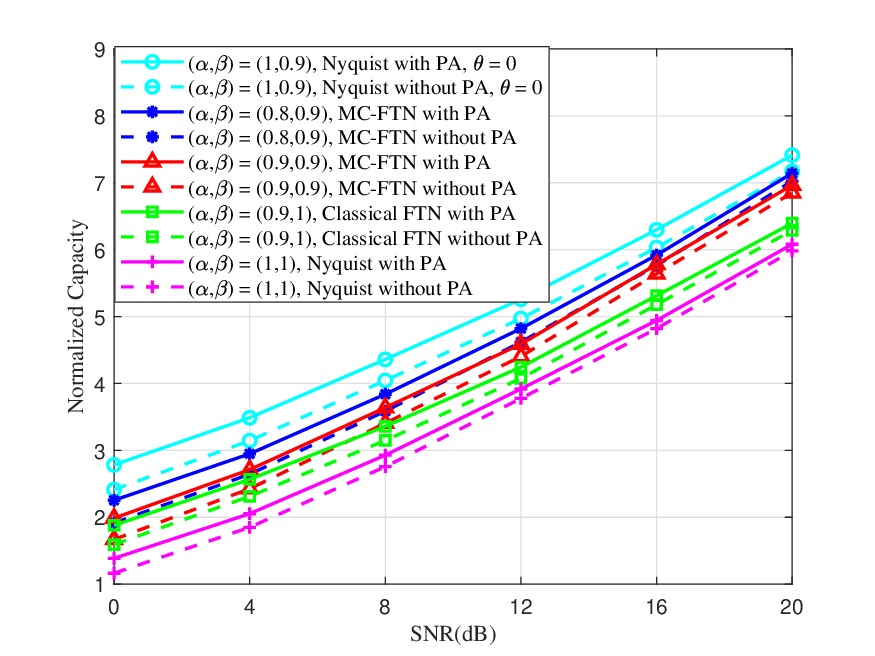}}
						\vspace{-2mm}
						\caption{\color{red} Normalized capacity $C^\mathrm{SISO}_\mathrm{nor}$ of our proposed MC-FTN-OTFS signaling scheme with and without PA. The compression factors are set as $(\alpha,\beta) = (1,1)$, $(\alpha,\beta) = (0.9,1)$, $(\alpha,\beta) = (0.9,0.9)$,  $(\alpha,\beta) = (0.8,0.9)$ and $(\alpha,\beta) = (1,0.9)$. }
						\label{Rab}
%						\vspace{-7mm}
					\end{figure}

					Fig. \ref{Rab} shows the normalized capacity performance of our proposed MC-FTN-OTFS signaling with parameters   $(\alpha,\beta) = (0.9,0.9)$ and  $(\alpha,\beta) = (0.8,0.9)$ with and without PA, while the parameters of the benchmarks are set as $(\alpha,\beta) = (1,1)$ and   $(\alpha,\beta) = (0.9,1)$ with  $\theta = 0.25$ for the RRC filter and $(\alpha,\beta) = (1,0.9)$ with $\theta=0$ for the ideal rectangular filter. The precoding matrix with PA is given by $\mathbf{P} = \mathbf{U}_{D} \boldsymbol{\Lambda}_P^{\frac{1}{2}}$, while the precoding matrix without PA is $\mathbf{P} = \mathbf{U}_{D}$, which is known as the conventional EVD-based precoding. From Fig. \ref{Rab}, we observe that the normalized capacity of MC-FTN-OTFS signaling with PA is higher than that of the counterparts without PA. Furthermore, our proposed MC-FTN-OTFS signaling outperforms the conventional FTN signaling~\cite{EVD,PRE3} and the Nyquist-based OTFS scheme~\cite{OTFS1} using the same RRC filter. Moreover, our proposed scheme with PA closely approaches the capacity performance of the rectangular-filter bound.
					%Additionally,  the normalized capacity increases as the compression factor $\alpha$ or $\beta$ decreases. 

					{\color{red}Fig. \ref{BERab} shows the BER performance of our proposed MC-FTN-OTFS signaling with $(\alpha,\beta)=(0.6,0.6)$ and $(\alpha,\beta)=(0.8,0.8)$, and the roll-off factor is set as $\theta = 0.75$.  The LDPC encoder and decoder are employed to correct  error bits and the minimum mean square error (MMSE) equalizer is adopted to recover the transmitted signals. For comparison, we also plot BER curves of the Nyquist-criterion-based signaling with $(\alpha,\beta)=(1,1)$, the classical FTN signaling with  $(\alpha,\beta)=(0.6,1)$ and  $(\alpha,\beta)=(0.8,1)$.}  From Fig. \ref{BERab}, we observe that as the compression factors $\alpha$ or $\beta$ decreases, the BER performance decreases moderately.  More specifically, our proposed signaling scheme can achieve a 	higher capacity at the cost of the moderate BER performance loss due to our proposed precoding scheme.

					%\begin{figure}[t]
					%	\center{\includegraphics[width=0.6\columnwidth]{BERa.eps}}
					%	\caption{The BER performance of our proposed FTN signaling for OTFS system. The roll-off factor of the RRC shaping filter is $\theta = 0.25$. The compression factors  are set to $\alpha = 1$, $\alpha = 0.9$ and $0.8$.}
					%	\label{BERa}
					%\end{figure}
					
					\begin{figure}[t]
						\center{\includegraphics[width=0.8\columnwidth]{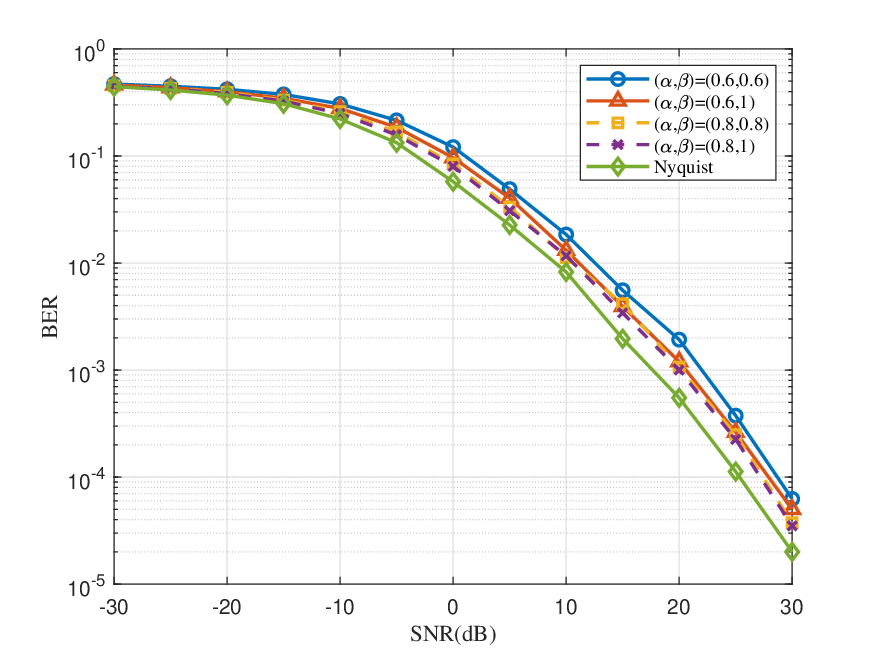}}\vspace{-2mm}
						\caption{{\color{red} BER performance of our proposed MC-FTN-OTFS signaling scheme under different compression factors, i.e., $(\alpha,\beta) = (0.6,0.6)$, $(\alpha,\beta) = (0.6,1)$, $(\alpha,\beta) = (0.8,0.8)$, $(\alpha,\beta) = (0.8,1)$,  and  $(\alpha,\beta) = (1,1)$.}}
						\label{BERab}
						\vspace{-2mm}
					\end{figure}

					\subsection{MC-FTN-OTFS Signaling Scheme for the MIMO System}
					In this subsection, we consider the performance of our proposed MC-FTN-OTFS signaling scheme for the MIMO system. Simulation results of capacity and BER of our proposed scheme are shown as follows.
					
					Fig. \ref{RMIMO} compares the normalized capacities of the MIMO system with different numbers of transmit and receive antennas.
					The numbers of transmit and receive antennas are set as $(N_T,N_R)=(1,1)$, $(N_T,N_R)=(2,2)$ and $(N_T,N_R)=(4,4)$. The compression factors are set as $(\alpha,\beta) = (0.9,0.9)$. We observe that the normalized capacity increases as the antenna numbers {increase}, owing to the spatial multiplexing gain. Moreover, the optimal WF-based precoding scheme and the low-complexity SIC-based precoding scheme with PA are also shown. We observe that the normalized capacity of the SIC-based precoder is modestly lower than that of the WF precoder and is almost identical at high SNR. Hence, the SIC-based precoder can achieve a close capacity performance and has lower complexity when the transmitted antenna number is large.
					
					\begin{figure}[]
						\center{\includegraphics[width=0.8\columnwidth]{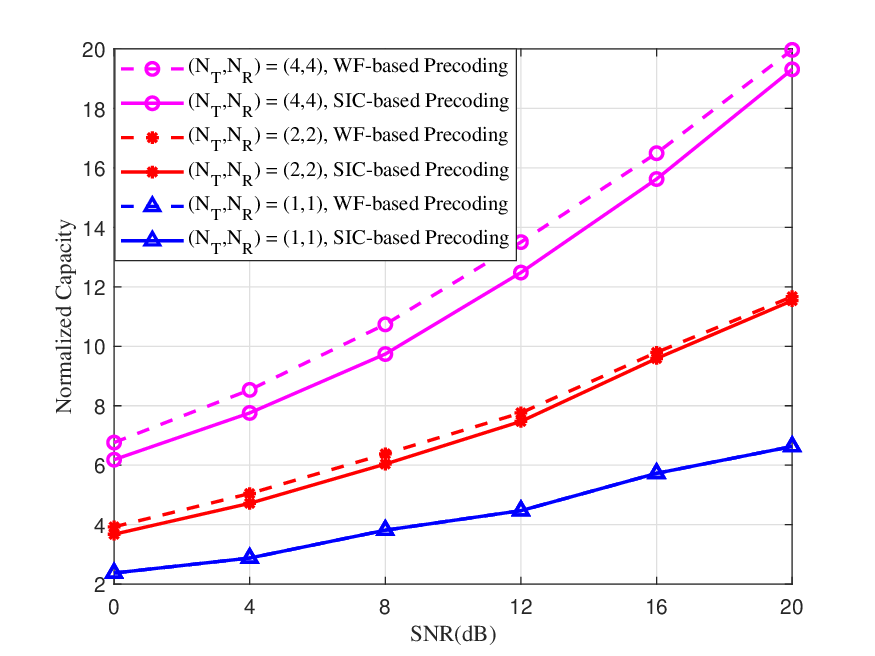}}
						\vspace{-2mm}
						\caption{\color{red} Normalized capacity $C^\mathrm{MIMO}_\mathrm{nor}$ of our proposed MC-FTN-OTFS signaling  with SIC-based and WF-based precoding schemes with different numbers of antennas, which are set as $(N_T,N_R)=(1,1)$, $(N_T,N_R)=(2,2)$ and $(N_T,N_R)=(4,4)$. The compression factors are set as $(\alpha,\beta) = (0.9,0.9)$.}
						\label{RMIMO}
%						\vspace{-7mm}
					\end{figure}
					
					\begin{figure}[t]
						\center{\includegraphics[width=0.8\columnwidth]{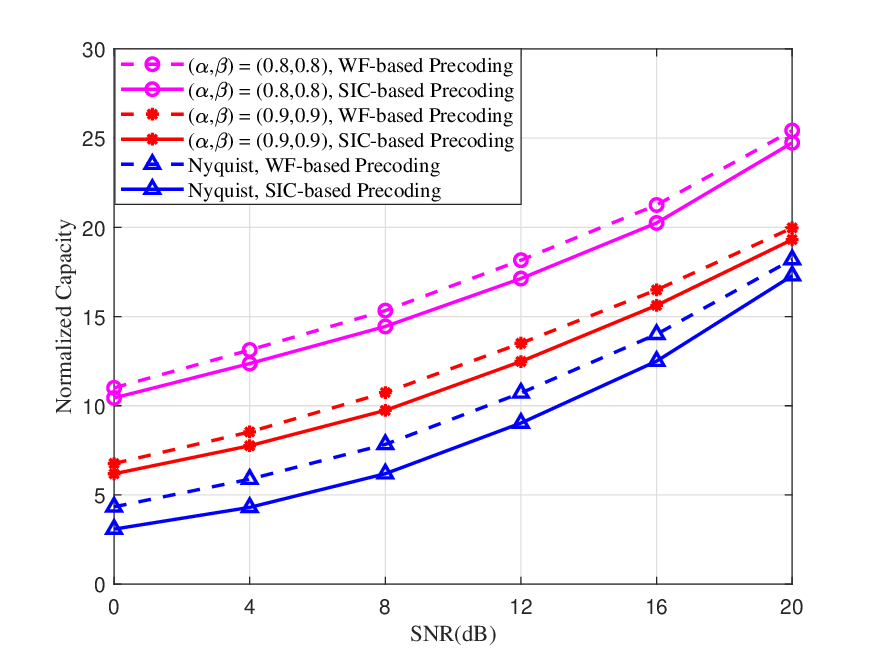}}
						\vspace{-2mm}
						\caption{{\color{red} Normalized capacity $C^\mathrm{MIMO}_\mathrm{nor}$ of our proposed MC-FTN-OTFS signaling  with SIC-based and WF-based precoding schemes under different compression factors, i.e., $(\alpha,\beta) = (1,1)$, $(\alpha,\beta) = (0.9,0.9)$ and  $(\alpha,\beta) = (0.8,0.8)$. The number of transmitted and received antennas are set as $(N_T,N_R)=(4,4)$.}}
						\label{RMIMO2}\vspace{-2mm}
					\end{figure}
					
					\begin{figure}[t]
						\center{\includegraphics[width=0.8\columnwidth]{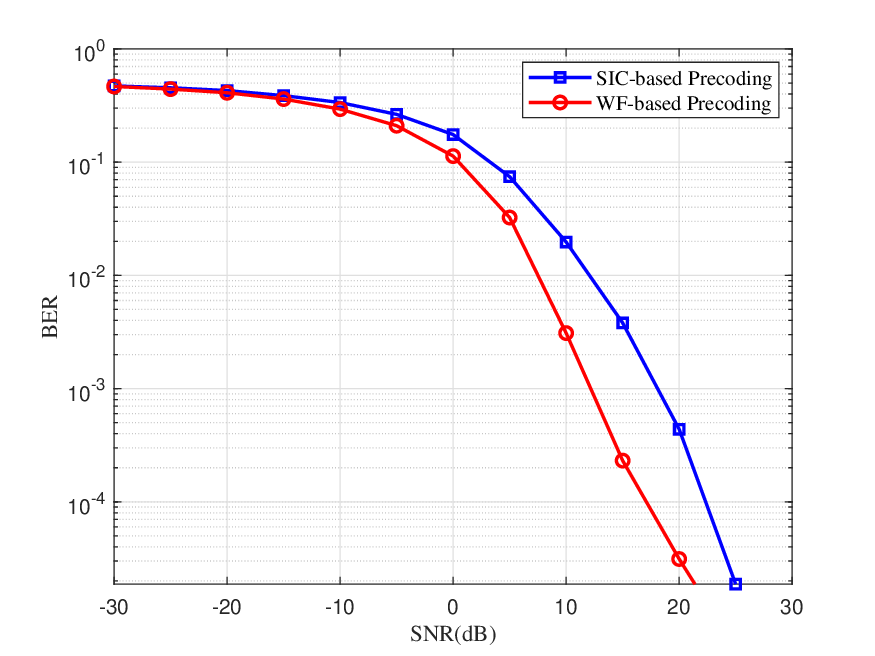}}\vspace{-2mm}
						\caption{{\color{red} BER performance of our proposed MC-FTN-OTFS signaling  with SIC-based and WF-based precoding schemes. The compression factors are set as $(\alpha,\beta) = (0.9,0.9)$. The numbers of transmitted and received antennas are set as $(N_T,N_R)=(4,4)$.}}
						\label{BERMIMO}
						\vspace{-7mm}
					\end{figure}

					Fig. \ref{RMIMO2} shows the normalized capacity of our proposed MC-FTN-OTFS signaling for the MIMO system with different compression factors  $(\alpha,\beta) = (0.9,0.9)$ and  $(\alpha,\beta) = (0.8,0.8)$. The normalized capacity curve of Nyquist-criterion-based signaling, i.e. $(\alpha,\beta) = (1,1)$, is also presented as the benchmark. We can clearly observe that as the compression factors $\alpha$ and $\beta$ decrease, the normalized capacity increases, exhibiting a similar trend as the SISO system. This is because the compression degree increases as the compression factors decrease, which finally achieves a higher capacity. Moreover, the  capacity performance of the SIC-based precoder is close to that of the optimal WF precoder with different compression factors, further demonstrating the effectiveness of our proposed SIC-based precoder.

					{\color{red} Fig. \ref{BERMIMO} shows the BER performance comparison of the proposed signaling schemes using the WF-based precoding and the low-complexity SIC-based precoding where the MMSE equalizer is applied at the receiver. Similar to Fig. \ref{BERab}, the LDPC encoder and decoder are employed. From Fig. \ref{BERMIMO}, we observe that the proposed low-complexity SIC-based precoder exhibits  a significantly lower computational
						complexity than  the WF precoder at  the  cost  of a  slight reduction in BER performance.}

					\section{Conclusions}\label{S6}
					In this paper, we proposed a novel EVD precoded MC-FTN-OTFS signaling scheme with optimal PA for time-varying channels. We first introduced the MC-FTN-OTFS modulation scheme for both SISO and MIMO systems and established the relationship for input and output signals. For SISO systems, we proposed an EVD precoding scheme with the optimal PA to eliminate the MC-FTN-induced ISI and ICI. For MIMO systems, we developed a low-complexity SIC-based precoder to avoid  high-dimensional calculations associated with the capacity maximization problem. Extensive simulation results demonstrate that our proposed EVD-precoded MC-FTN-OTFS signaling scheme can achieve a much larger capacity than the classic Nyquist-criterion-based signaling scheme. With the use of practical MMSE receiver, the degradation in BER performance is insignificant. Moreover, our proposed SIC-based precoder can effectively reduce the complexity for MIMO systems,  with a small loss of capacity compared to the optimal EVD precoder.

					\begin{appendices}
						\vspace{-1mm}
						\section{}
						
	We firstly calculate the continuous received signal $Y(f,t)$ after the MC-FTN modulation from (\ref{rt}), which can be calculated as (\ref{Yft}), which is shown at the top of the next page, 
	\begin{figure*}[!h]
			\vspace{-5mm}
		\centering
	\begin{align} \label{Yft}
	\begin{aligned}
	Y(f,t)   = & \int_{} g_{\mathrm{rx}}^{*}\left(t^{\prime}-t\right)\left[\int_{} \int_{} h(\tau, \nu) s\left(t^{\prime}-\tau\right) e^{j 2 \pi \nu\left(t^{\prime}-\tau\right)} d \tau d \nu+z(t')\right] e^{-j 2 \pi f\left(t^{\prime}-t\right)} d t^{\prime} \\
	= & \int_{} g_{\mathrm{rx}}^{*}\left(t^{\prime}-t\right)\left[h_i\sum_{n^{\prime}=0}^{N-1} \sum_{m^{\prime}=0}^{M-1} X_P\left[m^{\prime}, n^{\prime}\right] g_{\mathrm{tx}}\left(t^{\prime}-\tau_i-n^{\prime}\alpha T_0\right) e^{j 2 \pi m^{\prime}\beta \Delta f_0 \left(t^{\prime}-\tau_i-n^{\prime }\alpha T_0 \right)} \right. \\&\left. \quad \quad \quad e^{j 2 \pi \nu_i\left(t^{\prime}-\tau_i\right)} \right]  e^{-j 2 \pi f\left(t^{\prime}-t\right)} d t^{\prime}  +  \int_{} g_{\mathrm{rx}}^{*}\left(t^{\prime}-t\right)z(t')  e^{-j 2 \pi f\left(t^{\prime}-t\right)} d t^{\prime}  \\
	= & \sum_{n^{\prime}=0}^{N-1} \sum_{m^{\prime}=0}^{M-1} X_P\left[m^{\prime}, n^{\prime}\right] h_i \left\{\int_{} g_{\mathrm{rx}}^{*}\left(t^{\prime}-t\right) g_{\mathrm{tx}}\left(t^{\prime}-\tau_i-n^{\prime} \alpha T_0\right) e^{j 2 \pi m^{\prime} \beta \Delta f_0 \left(t^{\prime}-\tau_i-n^{\prime}\alpha T_0\right)} \right. \\&\left. \quad \quad \quad e^{j 2 \pi \nu_i \left(t^{\prime}-\tau_i \right)} e^{-j 2 \pi f\left(t^{\prime}-t\right)} d t^{\prime}\right\} +  Z(f,t), \end{aligned}
	\end{align}
	\hrulefill
	\vspace{-1mm}
	\end{figure*}
	where the transmit signal $s(t)$  and the delay-Doppler channel response $h(\tau,\nu)$ are respectively shown in (\ref{h-prior}) and (\ref{hdD}), the channel information term is defined as $H_{m,n}[m',n']=h_i \left\{\int_{} g_{\mathrm{rx}}^{*}\left(t^{\prime}-t\right)  g_{\mathrm{tx}}\left(t^{\prime}-\tau_i-n^{\prime} \alpha T_0\right)  e^{j 2 \pi m^{\prime}\! \beta \Delta f_0 \! \left(t^{\prime}-\tau_i-n^{\prime}\alpha T_0\! \right)} \right.$\\$\left. e^{j 2 \pi \nu_i \left(t^{\prime}-\tau_i \right)} e^{-j 2 \pi f\left(t^{\prime}-t\right)} d t^{\prime}\right\}$ and the noise term is defined as $Z(f,t) = \int_{} g_{\mathrm{rx}}^{*}\left(t^{\prime}-t\right)z(t')  e^{-j 2 \pi f\left(t^{\prime}-t\right)} d t^{\prime}$. 
						%We further define that the discrete noise $Z[m,n]=\left.Z(f, t)\right|_{f=m \beta \Delta f, t=n \alpha T_0}$. 
						Similar to \cite{Inteference}, we can further calculate the channel state information $H_{m,n}[m',n']$ in the MC-FTN-OTFS modulation as (\ref{H}), which is shown at the top of the next page.
				\begin{figure*}[!h]
					\centering
					\vspace{-5mm}
				\begin{align} \label{H}
				\begin{aligned}
				H_{m,n}\left[m^{\prime}, n^{\prime}\right] 
				=\sum_{i=1}^P h_i A_{g_{\mathrm{rx}}, g_{\mathrm{tx}}}\left(\left(m-m^{\prime}\right) \beta \Delta f_0-\nu_i,\left(n-n^{\prime}\right) \alpha T_0-\tau_i\right) \notag\\  e^{j 2 \pi\left(\nu_i+m^{\prime}\beta \Delta f_0 \right)\left(\left(n-n^{\prime}\right) \alpha T_0-\tau_i\right)} e^{j 2 \pi \nu_i n^{\prime} \alpha T} .
				\end{aligned}
				\end{align}	
				\hrulefill
				\vspace{-1mm}
				\end{figure*}
						Hence, the sampled received signal $Y[m,n]$ after sampling at $f=m\beta \Delta f_0$ and $t=n\alpha T_0$ is given by
						\begin{align}
						Y[m, n]=\sum_{m^{\prime}=0}^{N-1} \sum_{n^{\prime}=0}^{M-1} H_{m, n}\left[m^{\prime}, n^{\prime}\right] X_P\left[m^{\prime}, n^{\prime}\right]+Z[m, n],
						\end{align}
						where $Z[m,n]=\left.Z(f, t)\right|_{f=m \beta \Delta f_0, t=n \alpha T_0}$.
						
						\vspace{-1mm}
						\section{}

						Recalling (\ref{XTF}), (\ref{Ynm}) and (\ref{ykl}), we have (\ref{ykl2}) at the top of the next page
						\begin{figure*}
							\centering
							\vspace{-5mm}
								\begin{align}\label{ykl2}
							y[l,k]&=\frac{1}{\sqrt{N M}} \sum_{n=0}^{N-1} \sum_{m=0}^{M-1} \left[\sum_{m^{\prime}=0}^{N-1} \sum_{n^{\prime}=0}^{M-1} H_{m, n}\left[m^{\prime}, n^{\prime}\right] \left\{\frac{1}{\sqrt{N M}} \sum_{k'=0}^{N-1} \sum_{l'=0}^{M-1} x_{P}[l', k'] e^{j 2 \pi\left(\frac{n' k'}{N}-\frac{m' l'}{M}\right)} \right\} +Z[m, n]\right] e^{-j 2 \pi\left(\frac{n k}{N}-\frac{m l}{M}\right)} \notag \\
							&=\frac{1}{N M} \sum_{k^{\prime}=0}^{N-1} \sum_{l^{\prime}=0}^{M-1} h_{l,k}\left[l^{\prime}, k^{\prime}\right] x_P\left[l^{\prime}, k^{\prime}\right]+z[l,k],
							\end{align}
							\hrulefill
							\vspace{-1mm}
						\end{figure*}
					by defining 
						\begin{align}
						h_{l, k}\left[l^{\prime}, k^{\prime}\right]=\sum_{n=0}^{N-1} \sum_{m=0}^{M-1} \sum_{n^{\prime}=0}^{N-1} \sum_{m^{\prime}=0}^{M-1} H_{m, n}\left[m^{\prime}, n^{\prime}\right] e^{-j 2 \pi\left(\frac{n k}{N}-\frac{m \ell}{M}\right)} \notag \\ e^{j 2 \pi\left(\frac{n^{\prime} k^{\prime}}{N}-\frac{m^{\prime} e^{\prime}}{M}\right)}
						\end{align}
						and
						\begin{align}
						z[l, k]=\frac{1}{\sqrt{N M}} \sum_{n=0}^{N-1} \sum_{m=0}^{M-1} Z[m, n] e^{-j 2 \pi\left(\frac{n k}{N}-\frac{m l}{M}\right)}
						\end{align}
						
						Hence, we can write the DD domain input-output relationship in \emph{Proposition 2}, which completes the proof.
						
						\section{}
						We can rewrite (\ref{ynr}) as (\ref{ynr1}) at the top of the next page.
						\begin{figure*}
							\centering
							\vspace{-5mm}
							\begin{align} \label{ynr1}
							\mathbf{y}_{n_{r}}^{\mathrm{DD}}&=\sum_{n_{t}=1}^{N_{T}} \mathbf{H}_{n_{r}, n_{t}}^{\mathrm{DD}} \mathbf{P}_{n_t}\mathbf{x}^\mathrm{DD}_{n_t}+\mathbf{z}_{n_{r}}^{\mathrm{DD}}\notag \\
							&=\left[\mathbf{H}_{n_{r}, 1}^{\mathrm{DD}},\dots,\mathbf{H}_{n_{r}, N_T}^{\mathrm{DD}}\right]
							\left[\begin{array}{ccc}
							\mathbf{P}_{1} &   & \\
							& \ddots & \\
							&  &\mathbf{P}_{N_T}
							\end{array}\right]
							\left[\begin{array}{c}
							\mathbf{x}^\mathrm{DD}_{1} \\
							\vdots \\
							\mathbf{x}^\mathrm{DD}_{N_T}
							\end{array}\right]+\left[\begin{array}{c}
							\mathbf{z}^\mathrm{DD}_{1} \\
							\vdots \\
							\mathbf{z}^\mathrm{DD}_{N_T}
							\end{array}\right].
							\end{align}
							\hrulefill
						
						\end{figure*}

						Then we stack $N_R$ vectors $\mathbf{y}_{n_{r}}^{\mathrm{DD}}$ and obtain the MIMO received vector $\mathbf{y}_\mathrm{MIMO}^{\mathrm{DD}}$ as (\ref{yMIMO}),
						\begin{figure*}[!h]
							\centering
							\vspace{-5mm}
						\begin{align}\label{yMIMO}
						\mathbf{y}_\mathrm{MIMO}^{\mathrm{DD}}&=\left[\begin{array}{c}
						\mathbf{y}^\mathrm{DD}_{1} \\
						\vdots \\
						\mathbf{y}^\mathrm{DD}_{N_T}
						\end{array}\right] \notag \\
						&=\left[\begin{array}{cccc}
						\mathbf{H}^\mathrm{DD}_{1,1} & \mathbf{H}^\mathrm{DD}_{1,2} & \ldots &\mathbf{H}^\mathrm{DD}_{1,N_T}\\
						\vdots&\vdots& \ddots &\vdots \\
						\mathbf{H}^\mathrm{DD}_{N_R,1} & \mathbf{H}^\mathrm{DD}_{N_R,2} & \ldots &\mathbf{H}^\mathrm{DD}_{N_R,N_T}
						\end{array}\right]\left[\begin{array}{ccc}
						\mathbf{P}_{1} &   & \\
						& \ddots & \\
						&  &\mathbf{P}_{N_T}
						\end{array}\right]
						\left[\begin{array}{c}
						\mathbf{x}^\mathrm{DD}_{1} \\
						\vdots \\
						\mathbf{x}^\mathrm{DD}_{N_T}
						\end{array}\right]+\left[\begin{array}{c}
						\mathbf{z}^\mathrm{DD}_{1} \\
						\vdots \\
						\mathbf{z}^\mathrm{DD}_{N_T}
						\end{array}\right],
						\end{align}
						\hrulefill
						\vspace{-7mm}
						\end{figure*}
						where  $\mathbf{P}_\mathrm{MIMO}=\operatorname{diag}\left\{\mathbf{P}_{1}, \ldots, \mathbf{P}_{N_{T}}\right\} \in \mathbb{C}^{M N N_{R} \times M N N_{T}}$, we define $\mathbf{x}_\mathrm{MIMO}^\mathrm{DD}\in \mathbb{C}^{MNN_T\times 1}$ and $\mathbf{z}_\mathrm{MIMO}^\mathrm{DD}\in \mathbb{C}^{MNN_T\times 1}$ as the stacked vectors of $\mathbf{x}_\mathrm{n_r}^\mathrm{DD}$ and $\mathbf{z}_\mathrm{n_r}^\mathrm{DD}$. Hence, we can rewrite (\ref{yMIMO}) as $\mathbf{y}_{\mathrm{MIMO}}^{\mathrm{DD}}=\mathbf{H}_{\mathrm{MIMO}}^{\mathrm{DD}} \mathbf{P}_{\mathrm{MIMO}} \mathbf{x}_{\mathrm{MIMO}}^{\mathrm{DD}}+\mathbf{z}_{\mathrm{MIMO}}^{\mathrm{DD}}$, which completes the proof.

					\end{appendices}

					%&= \left [{\bf{X}}(0)^H,{\bf{X}}(1)^H, \ldots, {\bf{X}}(N-1)^H \right ]
					%\appendix
					%\section {}
					%
					%As discussed in Section III, the false angel is separated from the actual angle by $\frac{f_c}{f+f_c}$. So, to make the two indexes differ by $1$, we have
					%\begin{align}
					%\frac{{{f_c}}}{{{f_c} + {f_2}}} - \frac{{{f_c}}}{{{f_c} + {f_1}}} = \frac{1}{G},
					%\end{align}
					%where $f_1$ and $f_2$ denote the frequency of subcarriers. Then, $f_1$ can be expressed by $f_2$ as
					%\begin{align}
					%{f_1} = \frac{{{f_2}({f_c} + G{f_c}) + f_c^2}}{{(G - 1){f_c} + {f_2}}}.
					%\end{align}
					%
					%Then, the span between two subcarriers can be expressed as
					%\begin{align}
					%{f_1} - {f_2} = \frac{{{f_2}({f_c} + G{f_c}) + f_c^2 - {f_2}(G - 1) - {f_2}^2}}{{(G - 1){f_c} + {f_2}}},
					%\end{align}
					%where we define the frequency of subcarrier $f_2={\frac{{{n_p}W}}{{{N_p}}}}$

					\bibliographystyle{IEEEtran}
					\bibliography{IEEEabrv,Refference}
					
				\end{document}